\documentclass{aa} 
\usepackage[dvipsnames]{xcolor}
\usepackage[normalem]{ulem}
\usepackage{orcidlink}
\usepackage{subfigure}
\usepackage{siunitx}
\usepackage{CJKutf8} 
\usepackage{makecell}
\usepackage{graphicx}
\usepackage{txfonts}
\usepackage{amsmath}
\usepackage{hyperref}
\hypersetup{
     colorlinks=true,
     linkcolor=purple,
     citecolor=blue, 
     urlcolor=purple,
     }
\AtBeginDocument{\mathcode`v=\varv}
\newcommand\DM{\mathrm{DM}}

\newcommand\mus{\mu\mathrm{s}}

\DeclareFontFamily{U}{mathb}{\hyphenchar\font45}
\DeclareFontShape{U}{mathb}{m}{n}{
      <5> <6> <7> <8> <9> <10> gen * mathb
      <10.95> mathb10 <12> <14.4> <17.28> <20.74> <24.88> mathb12
      }{}
\DeclareSymbolFont{mathb}{U}{mathb}{m}{n}
\DeclareFontSubstitution{U}{mathb}{m}{n}

\let\dot\relax
\DeclareMathAccent{\dot}{0}{mathb}{"39}
\let\ddot\relax
\DeclareMathAccent{\ddot}{0}{mathb}{"3A}
\let\dddot\relax
\DeclareMathAccent{\dddot}{0}{mathb}{"3B}
\let\ddddot\relax
\DeclareMathAccent{\ddddot}{0}{mathb}{"3C}

\begin{document} 

   \title{Tackling artefacts in the timing of relativistic pulsar binaries: towards the SKA}
   \subtitle{}

   \author{Huanchen Hu \begin{CJK*}{UTF8}{gkai}(胡奂晨)\end{CJK*}\inst{1}\orcidlink{0000-0002-3407-8071}\thanks{huhu@mpifr-bonn.mpg.de}
          \and 
          Nataliya~K.~Porayko\inst{2,1} \orcidlink{0000-0002-6955-8040}\thanks{nporayko@mpifr-bonn.mpg.de}
          \and 
          Willem~van~Straten\inst{3}\orcidlink{0000-0003-2519-7375}
          \and
          Michael~Kramer\inst{1,4}\orcidlink{0000-0002-4175-2271}
          \and
          David~J.~Champion\inst{1}\orcidlink{0000-0003-1361-7723}   
          \and
          Michael~J.~Keith\inst{4}\orcidlink{0000-0001-5567-5492}
          }

   \authorrunning{H. Hu et al.}       
   \institute{Max-Planck-Institut f\"ur Radioastronomie, Auf dem H\"ugel 69, 53121 Bonn, Germany 
   \and Dipartimento di Fisica ``G. Occhialini'', Universita degli Studi di Milano-Bicocca, Piazza della Scienza 3, Milano, 20126, Italy
   \and Manly Astrophysics, 15/41-42 East Esplanade, Manly, NSW 2095, Australia
   \and Jodrell Bank Centre for Astrophysics, Department of Physics and Astronomy, University of Manchester, Manchester M13 9PL, UK
  }

   \date{Received MM DD, YYYY; accepted MM DD, YYYY}

  \abstract{Common signal-processing approximations produce artefacts when timing pulsars in relativistic binary systems, especially edge-on systems with tight orbits, such as the Double Pulsar. In this paper, we use extensive simulations to explore various patterns that arise from the inaccuracies of approximations made when correcting dispersion and Shapiro delay.
  In a relativistic binary, the velocity of the pulsar projected onto the line-of-sight varies significantly on short time scales, causing rapid changes in the apparent pulsar spin frequency, which is used to convert dispersive delays to pulsar rotational phase shifts.
  A well-known example of the consequences of this effect is the artificial variation of dispersion measure (DM) with binary phase, first observed in the Double Pulsar 20 years ago. 
  We show that ignoring the Doppler shift of the spin frequency when computing the dispersive phase shift exactly reproduces the shape and magnitude of the reported DM variations. 
  We also simulate and study two additional effects of much smaller magnitude, which are caused by the assumption that the spin frequency used to correct dispersion is constant over the duration of the sub-integration and over the observed bandwidth. 
  We show that failure to account for these two effects leads to orbital phase-dependent dispersive smearing that leads to apparent orbital DM variations. The functional form of the variation depends on the orbital eccentricity. 
  In addition, we find that a polynomial approximation of the timing model is unable to accurately describe the Shapiro delay of edge-on systems with orbits less than 4 hours, which poses problems for the measurements of timing parameters, most notably the Shapiro delay. 
  This will be a potential issue for sensitive facilities like the Five-hundred-meter Aperture Spherical Telescope (FAST) and the forthcoming Square Kilometre Array (SKA); therefore, a more accurate phase predictor is indispensable.
  }


   \maketitle


\section{Introduction}
\begin{figure*}[ht!]
    \centering
    \includegraphics[width=0.9\textwidth]{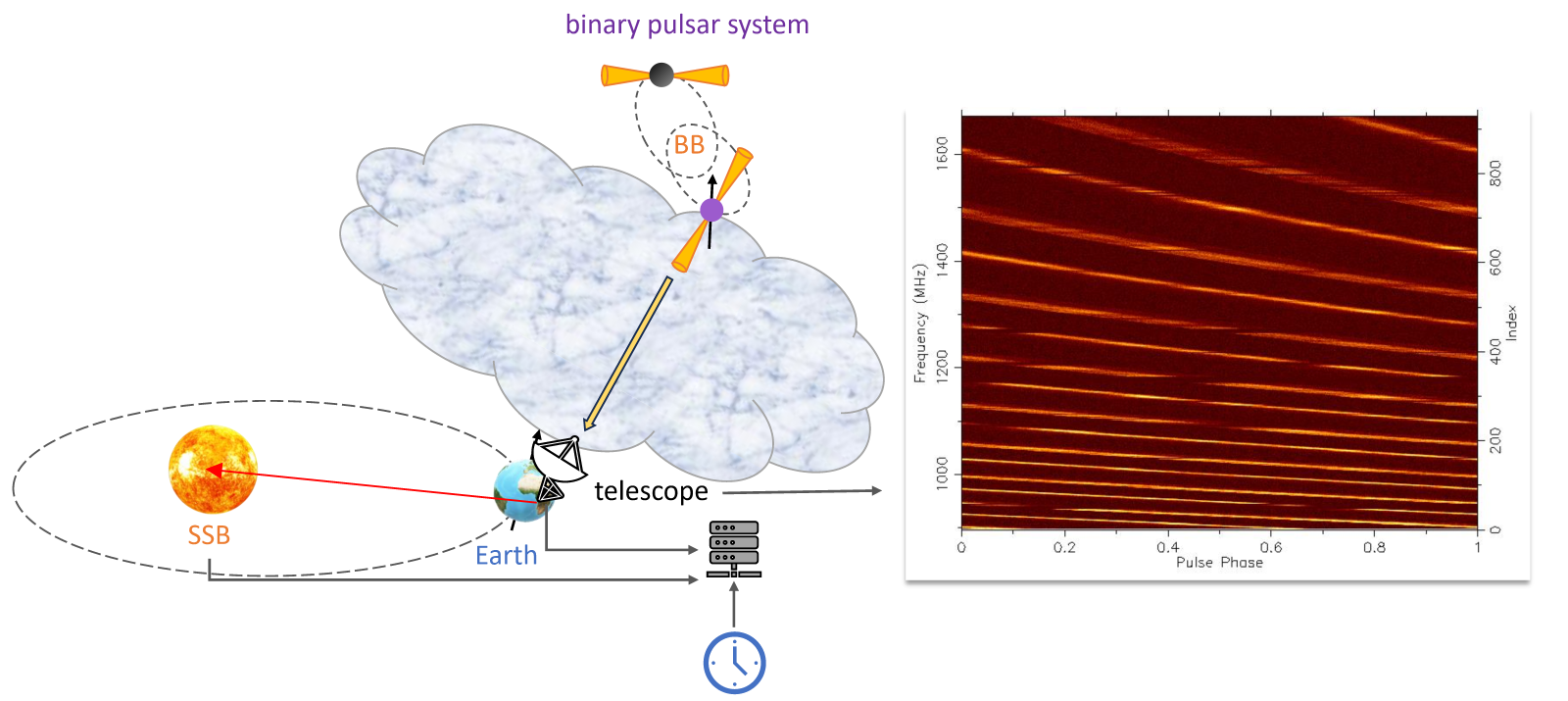}
    \caption{Illustration of the pulsar signal from a binary pulsar to timing data products. The binary motion of the pulsar around the binary barycentre (BB) causes Doppler shift of pulsar spin frequency. The pulsar signal propagates through ionised interstellar medium and suffers dispersive delay, causing the low-frequency signal to be delayed more than the signal at higher frequency. The plot on the right demonstrates this effect for PSR~J0737$-$3039A observed at the L band of the MeerKAT telescope \citep{Hu2023PhDT}. The received signal is further transferred to the Solar System Barycentre (SSB), where the Doppler shift of Earth's motion has to be corrected.}
    \label{fig:timing}
\end{figure*}

Relativistic binary pulsars play an important and sometimes unique role in the study of gravitational and fundamental physics.
Fifty years ago, in the summer of 1974, Russell Hulse and Joseph Taylor discovered the first binary pulsar \citep{HT1975ApJ}, taking tests of Einstein's general theory of relativity \citep[GR,][]{Einstein1915} beyond the weak-field slow-motion regime. 
The continued observations of the Hulse-Taylor pulsar enabled studies of strong-field and radiative aspects of the gravitational interaction, which provided the first evidence of the existence of gravitational waves \citep{Taylor1979Natur, Taylor1982ApJ, Taylor1989ApJ}.
Since then, more binary pulsars suitable for testing GR and alternative theories of gravity have been discovered \citep{Wex2020Univ, Freire2024LRR}. 
Relativistic effects in binaries are described by post-Keplerian parameters that can be precisely measured by pulsar timing. 
These parameters allow the determination of neutron star masses and even moments of inertia \citep{Hu+2020}, and as such plays an important role in constraining the equation of state of matter at supranuclear densities \citep{Oezel2016ARA&A, Hu2024Univ}. 
One particularly exceptional binary system for such studies is the Double Pulsar.

PSR J0737$-$3039A/B was discovered 21 years ago and remains the only Double Pulsar system \citep{Burgay+2003, Lyne+2004}. It consists of a 23-ms ``recycled'' pulsar A and a 2.8-s ``normal'' pulsar B, which orbit each other in a mildly eccentric ($e=0.088$) 2.45-h compact orbit. 
Its highly relativistic nature and nearly edge-on orientation make it an exceptional laboratory for testing GR effects and its higher-order terms in the orbital dynamics and photon propagation time. 
Various relativistic effects have been measured with extraordinary precision, including periastron precession, Einstein delay, and orbital period decay \citep{Kramer+2006Sci, Kramer+2021DP}, as well as relativistic spin precession of B \citep{Breton+2008Sci, Lower2024}, relativistic deformation of the orbit \citep{Kramer+2021DP}, and propagation effects of the signal from A through the strong gravitational field of B \citep{Kramer+2021DP,Hu+2022}. 
However, high precision measurement of the signal of interest can be complicated by a number of obstacles related to, e.g. ionised interstellar medium, strong profile evolution over the frequency band, or subtle effects that are not adequately handled by current data processing software. 
The latter is the main focus of this paper. 

A well-known example of inadequate data processing led to an apparent orbital variation in the dispersion measure (DM, the column density of free electrons along the line of sight to the pulsar) of PSR~J0737$-$3039A \citep[][later withdrawn]{Ransom+2004}.
Based on observations from the Green Bank Telescope (GBT), the amplitude of these variations was found to be as high as $\sim 0.05$~pc~cm$^{-3}$.
In the original article, the observed phase-dependent DM modulation was mainly attributed to propagation effects in the plasma of the interstellar medium and/or in a warm magnetised wind (presumably from pulsar B) that does not follow the standard cold-plasma dispersion relation. 
In later discourse, it was confirmed that the observed DM variations were spurious and due to incorrect data processing\footnote{See comments uploaded with \cite{Ransom+2004}.}. However, the origin, magnitude and functional form of the artefact have often been misunderstood. 
In this paper, we show that the apparent DM modulation is caused by completely neglecting the Doppler shift of the pulsar's spin frequency when computing the dispersive phase shift.
We discuss this ``Dispersive Doppler Variation'' (DDV) in detail in Section~\ref{sec:psr_doppler}.

DDV is sometimes confused with two second-order effects that are caused by the common practise of integrating the frequency-resolved average profile before correcting the dispersive delay.
First, when later computing and correcting the dispersive phase, it is necessary to assume a constant spin frequency over the sub-integration interval.
However, the pulsar spin frequency changes due to the Doppler effect, and neglecting this variation over the duration of the sub-integration results in orbital phase-dependent ``Temporal Dispersive Doppler Smearing'' (TDDS), as detailed in Section~\ref{sec:psr_temporal_smearing}.
Second, before correcting the dispersive delay, the signal received at higher radio frequency corresponds to later orbital phase (and, thus also, different spin frequency) compared to the signal received simultaneously at lower radio frequency.
If the differential spin frequency over the observed bandwidth is ignored when computing and correcting the dispersive phase, it leads to orbital phase-dependent
``Spectral Dispersive Doppler Smearing'' (SDDS).
In Section~\ref{sec:psr_spectral_smearing}, we show that this effect is generally much smaller than TDDS and becomes significant only at frequencies below 350~MHz.

Moreover, as already pointed out by \cite{Hu+2022}, folding with an incorrectly-configured phase predictor can lead to biases in the derived Shapiro delay parameters. 
This is because the polynomial approximation fails to fit the sharp Shapiro delay \citep{Shapiro1964}, which is described by a logarithmic function. 
Therefore, in this work, we also investigate the impact of this folding issue with regard to the compactness of the binary and the configuration of the predictor. 
As we will show in Section~\ref{sec:folding}, software typically used today can potentially limit timing precision and bias Shapiro delay measurements, especially for extremely sensitive instruments such as the Five-hundred-meter Aperture Spherical Telescope (FAST) and the Square Kilometre Array (SKA). Not only is this relevant for the Double Pulsar, but it is also important for other highly relativistic binary pulsars, such as PSR~J1946+2052 \citep{Stovall+2018ApJ} and PSR~J1757$-$1854 \citep{Cameron2018}, as well as for the upcoming discoveries in the near future. 

The aforementioned effects have been discussed in \cite{Hu+2022} and addressed when processing the MeerKAT data, but were not extensively explained.
Given that there is still a lack of literature on these effects, and to avoid further confusion in the community, we conduct an in-depth study of these artefacts with pulsar data simulations.
In the following, we give a brief introduction of pulsar timing and data processing software \textsc{psrchive} before a detailed description of the pulsar data simulator in Section~\ref{sec:simulator}. 
We reproduce the apparent DM variation caused by DDV and demonstrate two second-order effects, TDDS and SDDS, with simulations in Section~\ref{sec:binary_doppler} and present detailed mathematical expressions of these effects for the first time. 
Similar Doppler effects caused by the Earth's motion are discussed in Section~\ref{sec:topo_doppler}.
We demonstrate the folding issue with polynomial approximation in Section~\ref{sec:folding} with simulations and conclude the paper in Section~\ref{sec:summary} with an outlook into the future.

\begin{figure*}[ht!]
    \vspace{-10pt}
    \centering
    \includegraphics[width=\columnwidth]{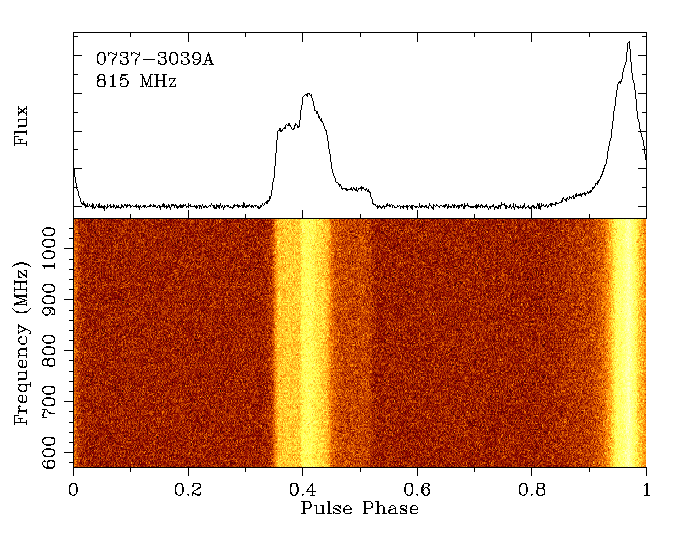}
    \includegraphics[width=\columnwidth]{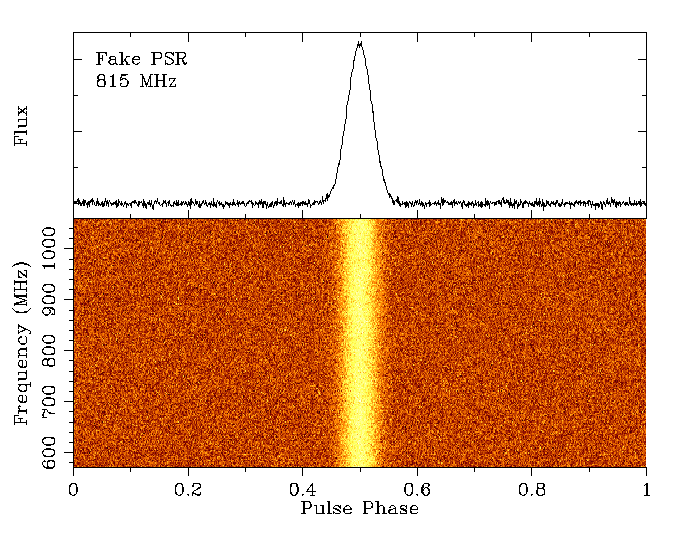}
    \vspace{-10pt}
    \caption{Example of simulated pulsar data files as shown by pazi command of the \textsc{psrchive} software: the Double Pulsar profile (left panel) and broad Gaussian profile used in the paper (right panel).}
    \label{fig:pazi_prof}
\end{figure*}
%
\section{Pulsar timing}
\label{sec:timing}
In this section we provide an overview of pulsar timing observation and data processing routines. A schematic diagram is shown in Fig.~\ref{fig:timing}. 
Pulsars are rotating neutron stars. As the pulsar rotates, a series of pulse signals from its magnetic poles travel through the ionised interstellar medium (IISM) before arriving at the radio telescope and being time-stamped by a maser clock.
Due to the interaction with the ionised media, the radio waves suffer a frequency-dependent time delay with respect to the signal at infinite frequency of
\begin{equation}
    t_\mathrm{DM} = \mathcal{D} \times \frac{\DM_0}{f^2} \,,
\end{equation}
where $\mathcal{D}$ is the dispersion constant defined as $1/(2.41000\times 10^{-4})\, \mathrm{MHz^{2}\,pc^{-1}cm^{3}s}$ in pulsar software \citep{MT1972} and $\DM_0 \simeq 48.9 \,\mathrm{pc \, cm^{-3}}$ is the intrinsic DM of PSR~J0737$-$3039A at the observing epoch. 
The dispersive time delay between two edge frequencies of the band $f_\mathrm{low}$ and $f_\mathrm{high}$ is therefore \citep{LK2004}:
\begin{align}
    \Delta t_\mathrm{DM} &= \mathcal{D}  \left(f_\mathrm{low}^{-2} - f_\mathrm{high}^{-2} \right) \times \DM_0 \,.
\end{align}
Since the time delay is inversely proportional to the square of the frequency, the effect is most prominent at lower frequencies, as illustrated in the plot on the right of Fig.~\ref{fig:timing}. 
After compensating for the dispersive delay (``de-dispersing''), one expects to see an aligned profile in frequency such as the left panel of Fig.~\ref{fig:pazi_prof}.

Single pulses are generally weak and vary in shape. To obtain a significant and stable pulse profile for pulsar timing, the incoming pulses are channelised in frequency and coherently de-dispersed within each channel, before being folded at the topocentric pulsar period over a sub-integration interval (typically $\sim 10$~s). 
Pulsars have excellent rotational stability. However, if the pulsar is moving in a binary system, the observed pulsar spin frequency $\nu$ will be Doppler shifted as a function of the orbital phase.
As we will show in this paper, the combination of dispersive delay and Doppler shift can introduce various artefacts if the data are not processed adequately.

Due to Earth's orbital and rotational motion in the Solar System, the time of arrival (TOA) of the pulsar signal received at the telescope (``topocentric'') must be transformed to the inertial reference frame at the Solar System Barycentre (SSB). Similar to the binary pulsar, Doppler frequency shift of Earth's motion has to be considered in addition to other barycentric corrections \citep[see e.g.][for a more complete overview]{LK2004}.

Pulsar observing systems that produce phase-resolved averages of observed quantities, such as the Stokes parameters, require a model that predicts the pulse phase as a function of time.  Evaluation of the physical timing model at every time sample is not computationally feasible; therefore, a polynomial is fit to the timing model and used to predict phase.
For example, the timing program {\sc tempo}\footnote{\url{https://tempo.sourceforge.net/}} \citep{TEMPO} produces one-dimensional polynomials
$\phi_{(1)}(t;f_0)$ that predict pulsar rotational phase as a function of topocentric time $t$ at a single topocentric reference frequency $f_0$.
In commonly-used signal processing software like \textsc{dspsr} \citep{vanStraten2011}\footnote{A digital signal processing software for filterbank pulsar data.}, the phase predicted by $\phi_{(1)}(t;f_0)$  is applied to all frequency channels that span the observed band. The dispersive delays between the frequency channels are corrected only after integrating over some period of time, $T$.
In this case, the dedispersed pulse phase at time $t$ and radio frequency $f$ is described by the following approximation, which can be compared to Eq.~(27) of \citet[][hereafter ``HEM2006'']{HEM2006}.
\begin{equation}
\tilde{\phi}(t,f) 
\simeq \phi_{(1)}(t;f_0) - \nu_{(1)}(t_\mathrm{mid};f_0) \Delta t_\mathrm{DM}(t_\mathrm{mid},f,f_0) 
\label{eq:phi_approx}
\end{equation}
Here, $\nu_{(1)} = d\phi_{(1)}/dt$ predicts the topocentric spin frequency of the pulsar at $t$ and $f_0$;
$t_\mathrm{mid}$ is an epoch near the middle of the sub-integration;
and $\Delta t_\mathrm{DM}$ corrects the dispersive delay between $f$ and $f_0$; this correction varies with time owing to the motion of the observatory with respect to the SSB.
Equation~(\ref{eq:phi_approx}) breaks down when the topocentric spin frequency of the pulsar varies significantly during either the sub-integration time $T$ or the dispersive delay interval $\Delta t_\mathrm{DM}$, whichever is greatest.
An exact expression for de-dispersed pulse phase is given by
\begin{align}
    \phi(t,f) = \phi_{(1)} (t - \Delta t_\mathrm{DM}[t, f, f_0]; f_0) \,,\label{eq:phi_exact}
\end{align}
Accurate calculation of the dispersive delay requires compensation of the Doppler shift due to motion of the observatory with respect to the SSB, and this physical model can also be approximated by a best-fit polynomial.
A similar approach is introduced by HEM2006, in which two-dimensional Chebyshev polynomials are used to predict
pulse phase as a function of topocentric time and frequency.
Commonly used signal processing software packages, like \textsc{dspsr}, \textsc{psrchive} and \textsc{presto}, do not yet fully 
implement the required polynomials, and continue to use the approximation presented in Eq.~\eqref{eq:phi_approx}.
The artefacts caused by the difference between $\phi(t,f)$ and $\tilde{\phi}(t,f)$ are discussed and demonstrated using simulations in Sections~\ref{sec:binary_doppler} and \ref{sec:topo_doppler}.

{\renewcommand{\arraystretch}{1.2}
\begin{table*}[ht]\small
\caption{Setup used to generate various datasets used in the current paper. }
\vspace{-15pt}
\centering
	\begin{tabular}{lccccc}
	\\
	\hline \hline
	Dataset & Central frequency & Bandwidth & Length of & Template & PSR ephemeris \\
  & (MHz) & (MHz) & sub-integration (s) &  & \\
	\hline
	Setting 1 & 452 & 100 & 10 &  J0737$-$3039A &  J0737$-$3039A\\ 
	\hline
	Setting 2a & 815& 490 & 10 & J0737$-$3039A/ & J0737$-$3039A/ \\ 
    & & & & broad Gaussian/ & eccentric ($e$=0.6)/\\ 
    & & & & narrow Gaussian & circular ($e$=0)\\ 
	\hline
    Setting 2b & 200 & 300 & 2 & J0737$-$3039A & circular ($e$=0) \\
    (SKA-Low) & & & & & \\
    \hline
    Setting 2c & 900 & 1100 & 10 & J0737$-$3039A & circular ($e$=0) \\
    (SKA-Mid)  & & & & & \\
	\hline
    Setting 3a & 815 & 490 & 10 &  J0737$-$3039A & J0737$-$3039A/\\
    & & & & & $P_\mathrm{b}=1$~h/\\
    & & & & & $P_\mathrm{b}=3$~h/\\
     & & & & & $P_\mathrm{b}=4.5$~h\\
	\hline
    Setting 3b & 7000 & 100 & 1 &  J0737$-$3039A & J0737$-$3039A/ \\ 
    & & & & & $P_\mathrm{b}=4.5$~h\\
	\hline
\end{tabular}
\tablefoot{Settings 1, 2a/b/c, and 3a/b are used to probe the pulsar orbital Doppler effect (Section~\ref{sec:psr_doppler}), orbital profile smearing effect (Section~\ref{sec:psr_temporal_smearing}), and folding issues with ``polyco'' (Section~\ref{sec:folding}), respectively. Number of frequency channels for all the setups is 200. For Settings 1, 2 and 3a the TZNSPAN parameter, which characterises the time resolution of the polynomial prediction, is fixed to 3 min. For Setting 3b TZNSPAN is 1 min. The template profile of J0737$-$3039A is constructed from the MeerKAT observations at L-band with a central frequency of 1283.582 MHz. Narrow and broad Gaussian profiles are simulated with width-period ratio (duty cycle) of 0.006 and 0.1, respectively. The parameter file of PSR J0737$-$3039A adopts measurements from \cite{Hu+2022}. 
To investigate how systematic error behaves under varying conditions, the PSR J0737$-$3039A ephemeris are modified as follows. Specifically, we alter the original orbital period ($P_\mathrm{b}=2.45$~h) and the eccentricity ($e=0.088$) for a series of tests with the values listed in the table.}
\label{table:detail}
\end{table*}
}
%
\section{Pulsar data simulator}
\label{sec:simulator}

In the current paper we make extensive use of the python-based \textsc{Pulsar Signal Simulator} (\textsc{PsrSigSim}) package, the main purpose of which is to produce realistic pulsar datasets. The details of the original version of the software developed by the North American Nanohertz Observatory for Gravitational Waves (NANOGrav) are described in \cite{Shapiro-Albert:2020tts}. We have kept the core of the native version, however, some of the modules have been modified to fit the purposes of our investigation\footnote{The original version of the code is publicly available through \url{https://github.com/PsrSigSim/PsrSigSim}, while the modified python library can be downloaded via \url{https://gitlab1.mpifr-bonn.mpg.de/nata/psrsigsim1_mpifr_gitlab.git}.}.
In particular, the pieces of the code have been updated to achieve phase connection with ``polyco'' and ``T2predict'' phase predictors, based on \textsc{tempo} and \textsc{tempo2} \citep{HEM2006} pulsar timing software, respectively. For the rest of the paper the phase-connection has been implemented with the ``polyco'' predictor, given that the complete timing model for the Double Pulsar is only available in \textsc{tempo}. See Section~\ref{sec:folding} for more details on phase connection.

The \textsc{PsrSigSim} has been used to create a series of phase-connected simulated pulsar observations. For this work, we have exclusively used the \textsc{filterbanksignal} class to produce channelised, time-resolved data cubes with pulsar observations folded modulo pulsar period. 
Thanks to built-in support of the \textsc{Pulsar Data Toolbox} \citep[\textsc{PDAT,}][]{pdat} the resulting data cubes are output in the \textsc{psrfits} format, which is compatible with the \textsc{psrchive} software \citep{Hotan+2004}, most commonly used to analyse pulsar astronomical data. 
\textsc{PsrSigSim} produces one pulsar data file per observational epoch with a customised number of frequency channels, phase bins and time sub-integrations. The profile shape and pulsar ephemeris are also fully user definable. Examples of generated pulsar data files are demonstrated in Fig.~\ref{fig:pazi_prof}. 
Although \textsc{PsrSigSim} software supports the creation of pulsar data files with frequency-evolving templates (through \textsc{portrait} class), in this work we ignore any possible profile shape change along the frequency band, both intrinsic and related to scattering in the IISM.
We note that in reality profile changes will further complicate the situation.

The functionality of \textsc{PsrSigSim} is particularly when investigating the influence of various frequency-dependent effects on pulsar timing. In particular, in this paper we simulate apparent chromatic artefacts induced by the motion of a pulsar in a binary system, such as those that are observed in the real data of PSR~J0737-3039A. To explore their nature and observational manifestation, a number of different pulsar ephemerides (e.g., with a pulsar in a circular or elliptical orbit) and profile shapes have been used. The setup has been adapted for each effect separately and is summarised in corresponding sections and Table~\ref{table:detail}.

\section{Doppler effects from binary pulsars}
\label{sec:binary_doppler}

\subsection{Dispersive Doppler Variation (DDV)}
\label{sec:psr_doppler}

In this section, we show that the effect apparently observed by \cite{Ransom+2004} in fact arises due to an unaccounted relativistic Doppler shift of pulsar spin frequency as the pulsar moves on its orbit in a binary system, which is practically equivalent to assuming a constant pulsar spin frequency over the orbit when computing the dispersive phase shift. 
The Doppler-shifted spin frequency observed at the observatory on Earth $\nu(t)$ is related to the emission frequency $\nu_0$ as 
\begin{equation}
    \nu(t) = \nu_0\,\sqrt{\frac{1-\beta(t)}{1+\beta(t)}} \,,
\end{equation}
where $\beta(t)$ is the ratio of the radial velocity of the pulsar $v_r^\mathrm{psr}(t)$ and the speed of light $c$. It is assumed that $\beta(t)= v_r^\mathrm{psr}(t)/c>0$ if the pulsar is moving away from us.
The pulse phase difference from dispersive delay $\Delta \tilde{\phi}$ is then given by the product of the dispersive delay time $\Delta t_\mathrm{DM}$ and the observed spin frequency $\nu(t)$:
\begin{align}
    \Delta \tilde{\phi} &= \mathcal{D}  \left(f_\mathrm{low}^{-2} - f_\mathrm{high}^{-2} \right) \times \DM_0  \times \nu_0  \sqrt{\frac{1-\beta(t)}{1+\beta(t)}} \,.
    \label{eq:bnry_dm}
\end{align}
However, if the data are de-dispersed using $\nu_0$ (instead of $\nu_\textrm{obs}$) as if the pulsar were fixed at the binary barycentre (as if ``solitary''), this would then lead to an apparent DM of $\DM' = \DM_0 + \delta\DM$, and pulse phase difference
\begin{equation}
    \Delta \phi_\mathrm{solitary} = \mathcal{D}  \left(f_\mathrm{low}^{-2} - f_\mathrm{high}^{-2} \right) \times \DM' \times \nu_0 \,.
    \label{eq:nobnry_dm}
\end{equation}
If the phase shift caused by dispersion (Eq.~\ref{eq:bnry_dm}) is corrected without accounting for the Doppler shift of spin frequency (Eq.~\ref{eq:nobnry_dm}), by equalling Eqs.~\eqref{eq:bnry_dm} and \eqref{eq:nobnry_dm}, one obtains the apparent DM variation
\begin{align}
    \delta\DM &= \left(\sqrt{\frac{1-\beta(t)}{1+\beta(t)}}-1 \right)\, \DM_0\,.
\end{align}
Using the Taylor expansion, we can simplify it to first order as
\begin{align}
    \delta\DM &\approx -\beta(t) \times \DM_0 = -\frac{v_r^\mathrm{psr}(t)}{c} \, \DM_0 \,.\label{eq:deltaDM}
\end{align}

For PSR~J0737$-$3039A, the orbital velocity is about a thousandth of the speed of light, given that the orbit is nearly edge on. The maximum DM variation is thus $\delta\DM_\mathrm{max} \simeq 10^{-3} \, \DM_0 = 0.05 \,\mathrm{pc \,cm^{-3}}$. 
This is consistent with the DM amplitude reported in \cite{Ransom+2004}. 
The occurrence of this error can also be expressed following Eq.~\eqref{eq:phi_approx} as:
\begin{align}
    \phi_1 (t,f) = \phi (t, f_0) -  \nu_0 \,\Delta t_\mathrm{DM} (f, f_0) \,.
\end{align}

\begin{figure}[t]
    \centering
    \includegraphics[width=\columnwidth]{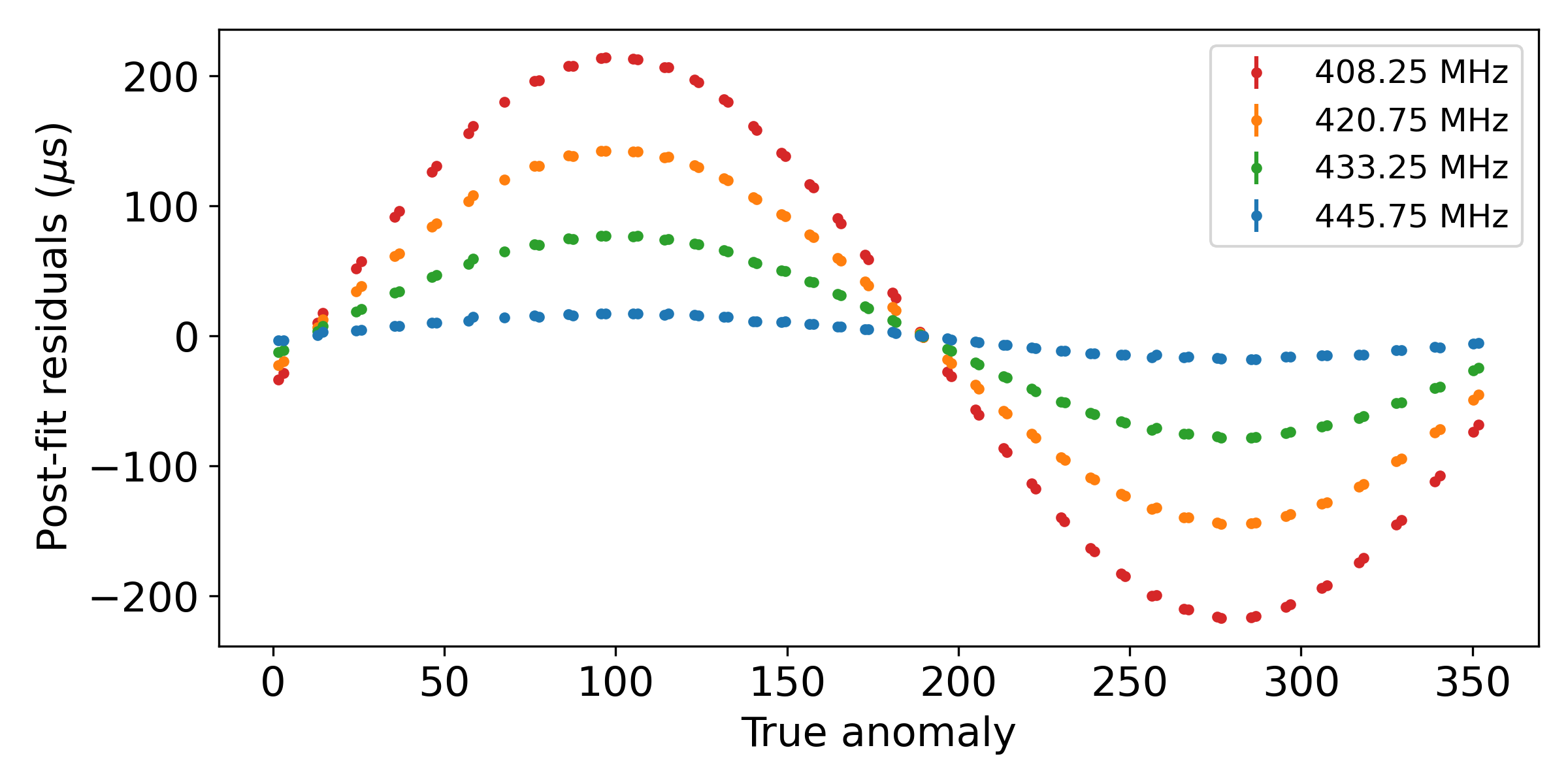}
    \vspace{-15pt}
    \caption{Timing variations of simulated by \textsc{PsrSigSim} software as a function of true anomaly for 4$\times$12.5 MHz sub-bands covering a 50 MHz band centered at 427 MHz. The setup repeats the one used in the \cite{Ransom+2004}. The apparent variations in timing residuals have the same magnitude of $\sim 200\,\mu$s as reported by \cite{Ransom+2004}. }
    \label{fig:res_doppler}
    \vspace{-5pt}
\end{figure}

\begin{figure}[t]
    \vspace{-5pt}
    \centering
    \includegraphics[width=\columnwidth]{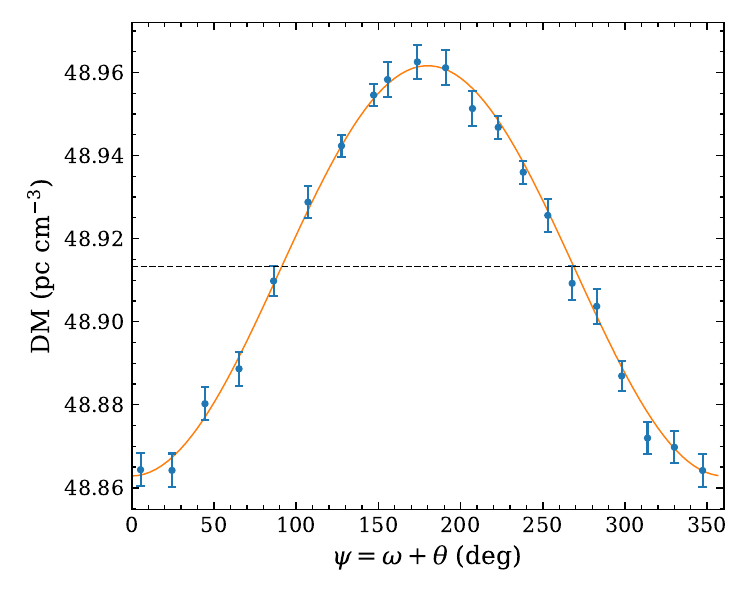}
    \vspace{-20pt}
    \caption{Reproduced apparent DM variation by de-dispersing the simulated data with a non-binary model (shown as blue points), i.e., ignoring the Dispersive Doppler Variation (DDV). The data are plotted against pulsar's angular orbital position $\psi=\omega+\theta$ ($\omega$ is the longitude of periastron and $\theta$ is the relativistic true anomaly) with respect to the ascending node. Both amplitude and phase are consistent with Fig.~2 in \cite{Ransom+2004}, given that $\omega \sim 80\si{\degree}$. The dashed horizontal line indicates the DM used in the simulation, and the orange line represents the calculation with Eq.~\eqref{eq:deltaDM}, which matches perfectly with the simulation.
    }
    \label{fig:dm_doppler}
\end{figure}

\begin{figure}[t]
    \centering
    \includegraphics[width=\columnwidth]{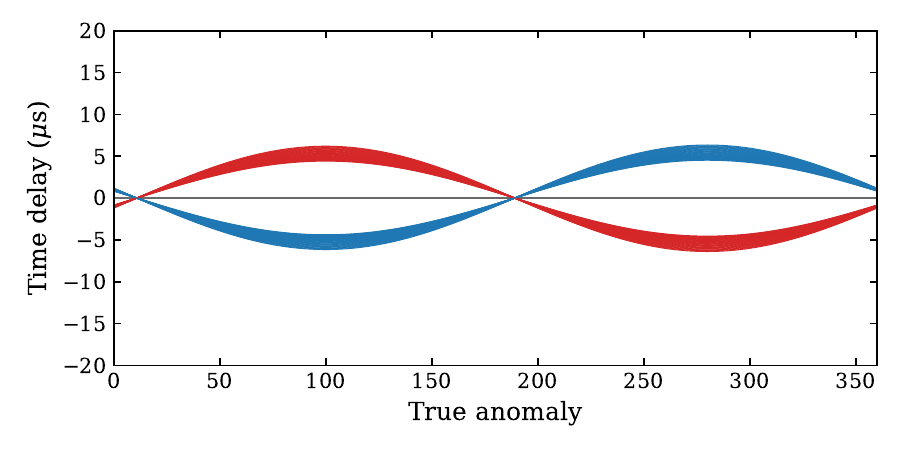}
    \vspace{-20pt}
    \caption{Dispersive time delay within a narrow 2~MHz-band as a comparison to Fig.~\ref{fig:res_doppler}. For all 25 sub-bands, the time difference between the lower edge of the sub-band and the central frequency is shown in red, while the time difference between the upper edge and central frequency is shown in blue. The lower the frequency of the sub-band, the greater the amplitude of the time delay. As one data point is generated per sub-band for a given sub-integration, this adds a $\sim 10\,\mus$ error to the data.}
    \label{fig:narrowband}
\end{figure}

In order to reproduce the observed behaviour shown in \cite{Ransom+2004}, we simulated pulsar data files using the \textsc{PsrSigSim} software with the following setup. The profile and ephemeris are those of the Double Pulsar. 
The generated 4-minute data files have a time resolution of 10 seconds. The total bandwidth of 100 MHz is split into 200 channels with a central frequency of 452 MHz. Details of the simulation correspond to the Setting~1 described in Table~\ref{table:detail}.
The pulsar software package \textsc{psrchive} corrects the frequency-dependent effects in relation to the centre frequency, while the version of Berkeley-Caltech Pulsar Machine \citep[BCPM,][]{2004ApJ...613L.137K} used the highest frequency of the band as the reference value. 
Therefore, the total bandwidth of simulated data files is twice the bandwidth of the original observations. 
The upper part of the frequency was discarded later in the post-processing. The simulated data files were de-dispersed with a modified parameter file that did not contain the binary model (in other words we assumed that the spin frequency is constant over the binary orbit), which effectively introduce the effect of uncorrected orbital Doppler shift of the spin frequency. All other processing steps were performed with the re-installed original parameter file of J0737$-$3039A. The results are shown in Figs.~\ref{fig:res_doppler} and \ref{fig:dm_doppler}. Fig.~\ref{fig:res_doppler} presents the timing residuals in four different sub-bands as a result of the uncorrected Doppler effect, while the apparent DM measured from the simulated data is demonstrated in Fig.~\ref{fig:dm_doppler}. 

Therefore, as confirmed by the mathematics and further with simulations, the incomplete correction of the Doppler shift of the pulsar spin frequency, combined with the dispersive delay caused by the IISM, creates apparent DM variations of the required magnitude.

In the past, this problem has been diminished by splitting the timing data into narrow sub-bands and folding each sub-band using a one dimensional phase predictor computed with a reference frequency equal to the centre of the band. 
To demonstrate this, we divide these data (402-452~MHz) into 25 sub-bands, each with a bandwidth of 2~MHz, and calculate the time delays within the sub-bands using Eq.~\eqref{eq:bnry_dm}. The results are shown in Fig.~\ref{fig:narrowband}, where the difference between the lower (upper) edge and the centre of the sub-bands is shown in red (blue) for all 25 sub-bands.
The maximum amplitude corresponds to the sub-band with the lowest frequency.
The error introduced by this effect is reduced from $\sim 200\,\mus$ in Fig.~\ref{fig:res_doppler} to a level of $\sim 5\,\mus$, which is well below the observational error of the pulse arrival time ($\sim 20\,\mus$ for the GBT data), and therefore has little impact on the timing precision. 
The bandwidth can be chosen in such a way that the time difference within the sub-bands is smaller than the observational error. 
This approach was employed in the analysis of the 16-yr data \citep{Kramer+2006Sci, Kramer+2021DP}, which helped to improve the timing precision before \textsc{psrchive} became available (while minimising the higher-order spectral smearing effect discussed in the following section). 
As \textsc{psrchive} already takes into account the Doppler effect of spin frequency properly, such apparent orbital DM variation should no longer be present. However, the effect may still occur when the data are processed by other software such as \textsc{sigproc}.

\subsection{Second-order Dispersive Doppler effects}
\label{sec:doppler_smearing_effects}

In addition to the first-order dispersive Doppler effect, DDV, two second-order effects of much smaller magnitude occur with the current data processing software and cause profile smearing.
The magnitude of these effects can be estimated using the frequency-dependent difference between $\tilde\phi$ and $\phi$ (Eqs.~\ref{eq:phi_approx} and~\ref{eq:phi_exact}, respectively), which is approximated using the Taylor expansion of the phase predictor,
\begin{equation}
\phi_{(1)}(t;f_0) = \phi_0 + \dot\phi_0 \Delta t + \frac{1}{2} \ddot\phi_0 \Delta t^2 + \frac{1}{6} \dddot\phi_0 \Delta t^3 + \ldots \; ,
\end{equation}
where $\Delta t = t - t_\mathrm{mid}$,
such that the topocentric spin frequency is
\begin{equation}
\nu_{(1)}(t;f_0) = \dot\phi_0 + \ddot\phi_0 \Delta t + \frac{1}{2} \dddot\phi_0 \Delta t^2 + \ldots \;.
\end{equation}
Ignoring any variation in the motion of the observatory during the sub-integration time, all dispersive Doppler smearing artefacts caused by the motion of the pulsar are described by the phase difference,
\begin{equation}
\begin{aligned}
\label{eq:delta_phi}
\Delta\phi(t,f) =&\; \tilde\phi(t,f) - \phi(t,f)  \\ 
%
= &\; \Delta t_\mathrm{DM}(t,f) \left[ \ddot\phi_0  \Delta t 
 - \frac{1}{2} \dddot\phi_0 \Delta t^2 + \ldots \right] \\ 
%
 & - \frac{1}{2} \Delta t_\mathrm{DM}^2(t,f) \left[ \ddot\phi_0 + \dddot\phi \Delta t + \ldots \right] \\
 & + \frac{1}{6} \dddot\phi_0 \Delta t_\mathrm{DM}^3(t,f) + \ldots 
\end{aligned}
\end{equation}
The leading term in this equation (proportional to $\Delta t_\mathrm{DM}$) describes the Temporal Dispersive Doppler Smearing (TDDS) effect, while
the second term represents the Spectral Dispersive Doppler Smearing (SDDS) effect. 
The magnitude and manifestation of these effects in pulsar timing data are discussed in the following sections.

\subsubsection{Temporal Dispersive Doppler Smearing (TDDS)}
\label{sec:psr_temporal_smearing}
\begin{figure}[t]
    \centering
    \includegraphics[width=\columnwidth]{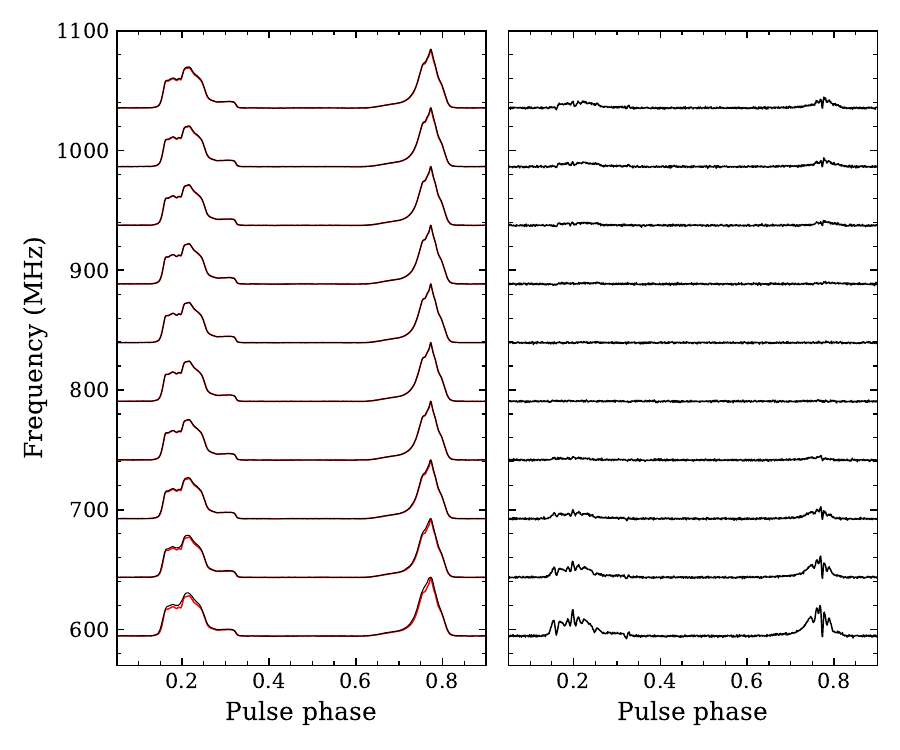}
    \vspace{-20pt}
    \caption{Profile comparison using the J0737$-$3039A profile with a circular orbit. The black lines in the left panel show the profile of processed data at different frequencies as compared to the standard template in red, whereas the right panel shows the difference between the smeared profile and the standard profile (magnified by five times) without any phase shift.}
    \label{fig:0737_cir_profile}
    \vspace{-10pt}
\end{figure}
\begin{figure}[t]
    \centering
    \includegraphics[width=\columnwidth]{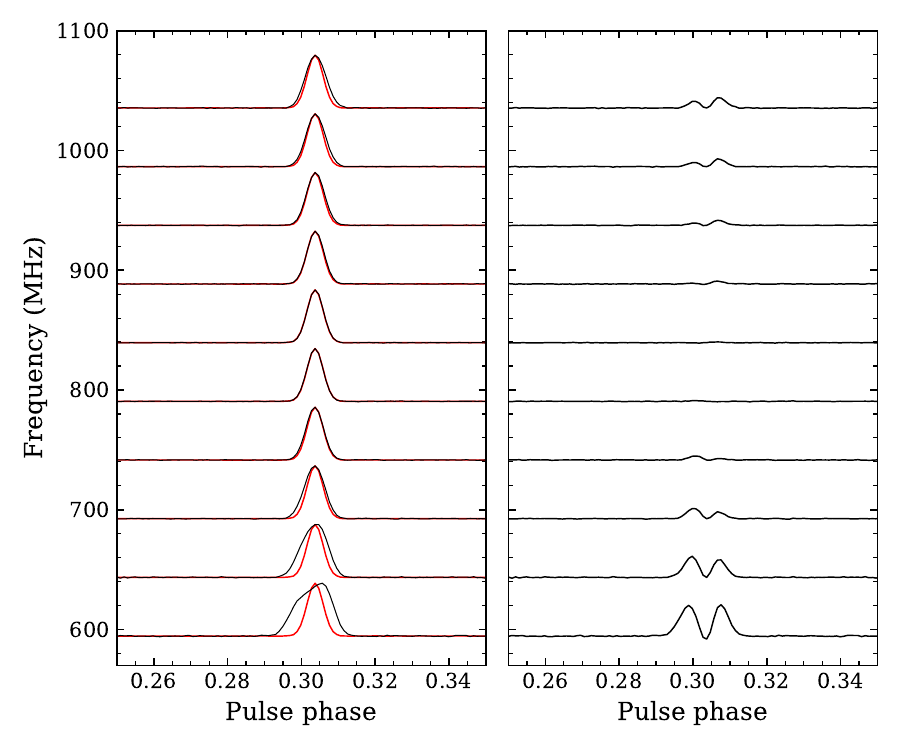}
    \vspace{-10pt}
    \caption{Same as Fig.~\ref{fig:0737_cir_profile} but for a narrow Gaussian profile with a circular orbit. Note that the pulse phase is zoomed to show the details of pulse profile and that the residuals in the right panel are not magnified.}
    \label{fig:0737_cir_narrowgaus}
    \vspace{-5pt}
\end{figure}

TDDS arises when the temporal variation of the Doppler-shifted pulsar spin frequency is neglected in the calculation and correction of the dispersive phase shift.  This occurs whenever the average profile is integrated before the dispersive delay is corrected, and the amount of smearing increases with integration length.  

At conjunction, where $\dddot\phi_0 = 0$, the total dispersive smearing over the sub-integration time $T$ is approximately
\begin{align}
\delta \phi_2(f,T) 
\simeq & \; \Delta\phi(t+T/2,f) - \Delta\phi(t-T/2,f) \nonumber \\
= & \; \ddot\phi_0 \Delta t_\mathrm{DM}(f,f_0) T \nonumber \\
= & \; \frac{a_r}{c} \nu_0 \,\Delta t_\mathrm{DM}(f,f_0) T \;, 
\label{eq:temporal_max}
\end{align}
where
$\nu_0 = \nu_{(1)}(t_\mathrm{mid},f_0)$ is the pulsar spin frequency at the mid-time of the sub-integration,
$a_r$ is the radial acceleration of the pulsar, and
$c$ is the speed of light.
Although maximal at conjunction, this smearing is also symmetric for either a circular orbit or an elliptical orbit for which periastron is at conjunction
(for the first half of $T$, $\Delta t < 0$, and for the second half,
$\Delta t > 0$).
Therefore, in such cases, the impact on arrival times is minimal at conjunction.
In contrast, at the (ascending and descending) nodes, where $\ddot\phi_0 = 0$, TDDS is dominated
by the $\dddot\phi_0$ term and varies quadratically with $\Delta t$.  Therefore, the smearing is asymmetric and the impact on arrival times is maximal at the nodes.

Predicting the impact of smearing on arrival times is  not trivial due to its strong dependence on the intrinsic shape of the average pulse profile. 
However, estimating the amount of smearing still provides a useful order of magnitude estimate.
After 5 minutes of integration using a polynomial computed for $f_0=815$~MHz to fold a signal at $f=570$~MHz, there is $\sim 70\, \mu$s of smearing in PSR~J0737-3039A. 
This second-order smearing is eliminated by correcting the dispersive delay before any temporal-averaging is applied.
\begin{figure}[t]
    \centering
    \includegraphics[width=\columnwidth]{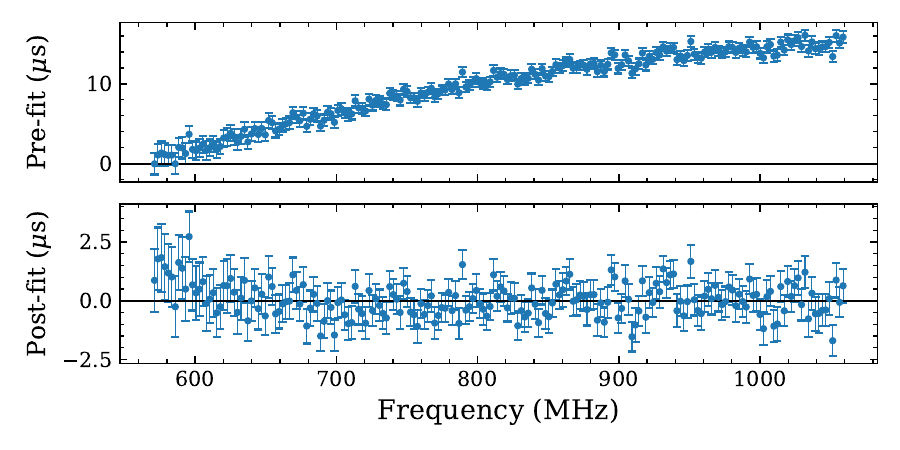} 
    (a)
    \includegraphics[width=\columnwidth]{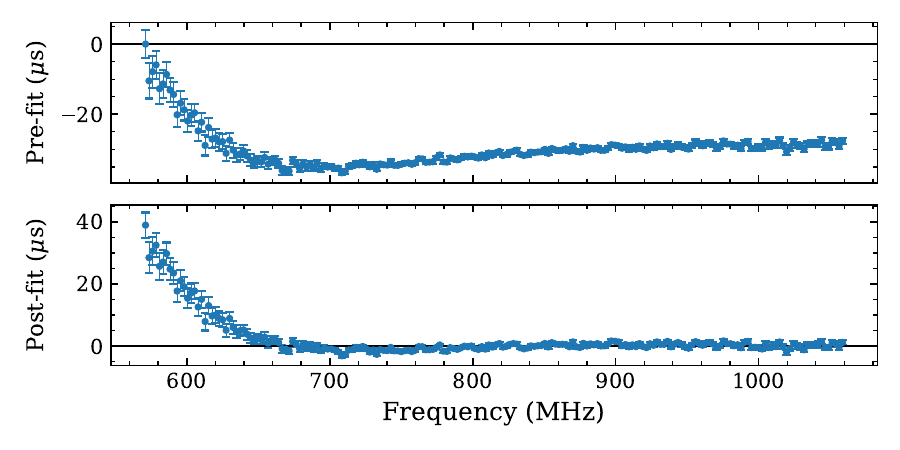}
    (b)
    \caption{Pre-fit and post-fit residuals after fitting for DM are shown in relation to frequency. Both figures are for the same orbital phase ($\psi\sim 130\si{\degree}$) but with different pulse profiles. (a) Circular orbit with the J0737$-$3039A profile (see Fig.~\ref{fig:0737_cir_profile}), residuals can mostly be absorbed by fitting DM. (b) Circular orbit with a narrow Gaussian profile (see Fig.~\ref{fig:0737_cir_narrowgaus}), only residuals $\gtrsim$660~MHz are absorbed by fitting for the DM. }
    \label{fig:cir_dp_residual}
\end{figure}

To demonstrate this effect, we simulate pulsar data files at frequencies similar to the UHF band of MeerKAT (570-1060~MHz) with different profiles and orbital eccentricities. 
The specification of the simulated dataset is described by Setting 2a in Table~\ref{table:detail}. The data were folded every 30 minutes without prior de-dispersion, which would otherwise have mitigated the effect. 
Figs.~\ref{fig:0737_cir_profile} and \ref{fig:0737_cir_narrowgaus} show how the effect alters the PSR~J0737$-$3039A profile and a narrow Gaussian profile (width-period ratio of 0.006) at a chosen orbital phase near the superior conjunction (ranging from $88\si{\degree}$ to $160\si{\degree}$) where the smearing effect is prominent. A circular orbit is assumed in both cases. 
It can be seen that after folding the pulse profiles (black lines in the left panels) are broadened at lower and upper parts of the band, compared to the standard template in red. 
The difference in the black and red profiles is shown on the right panels. 

For each of these profile shapes, Fig.~\ref{fig:cir_dp_residual} shows the arrival-time residuals after fitting for DM. 
For a relatively wide profile such as that of J0737$-$3039A, the apparent DM can be effectively removed by the fit. As a result, nearly flat residuals as a function of frequency are observed in Fig.~\ref{fig:cir_dp_residual}~(a). 
Similar results are obtained in a simulation with a broad Gaussian profile (width-period ratio of 0.1). 
This suggests that for wide pulse profiles, TDDS leads to residuals that can be reabsorbed with an inverse quadratic dependence on the observation frequency, which results in biased DM estimates. 

However, for narrow profiles, the residuals have a different functional form, as is illustrated in Fig.~\ref{fig:cir_dp_residual}~(b). In contrast to the broad profile case, significant spectral structure remains in the residuals after fitting for DM.
From Fig.~\ref{fig:0737_cir_narrowgaus} one can see that the profile is predominantly smeared to the left (right) at the lower (upper) part of the frequency band. Therefore, it is reasonable to expect that the residuals should follow the same "DM-like" trend as seen in Fig.~\ref{fig:cir_dp_residual}(a). This is indeed the case for residuals $\gtrsim$660~MHz. However, below $\sim$660~MHz, although the smearing is wider on the left, it has a higher peak on the right (see Fig.~\ref{fig:0737_cir_narrowgaus}). This asymmetric smearing causes the template to be matched to the peak component on the right-hand side of the smeared profile, corresponding to positive residual delay. This anomaly is not observed when the width of the pulse is substantially greater than the timescale of the TDDS. Despite this anomaly, the resulting apparent DM variations are similar for different pulse widths.

\begin{figure}[t]
    \centering
    \includegraphics[width=\columnwidth]{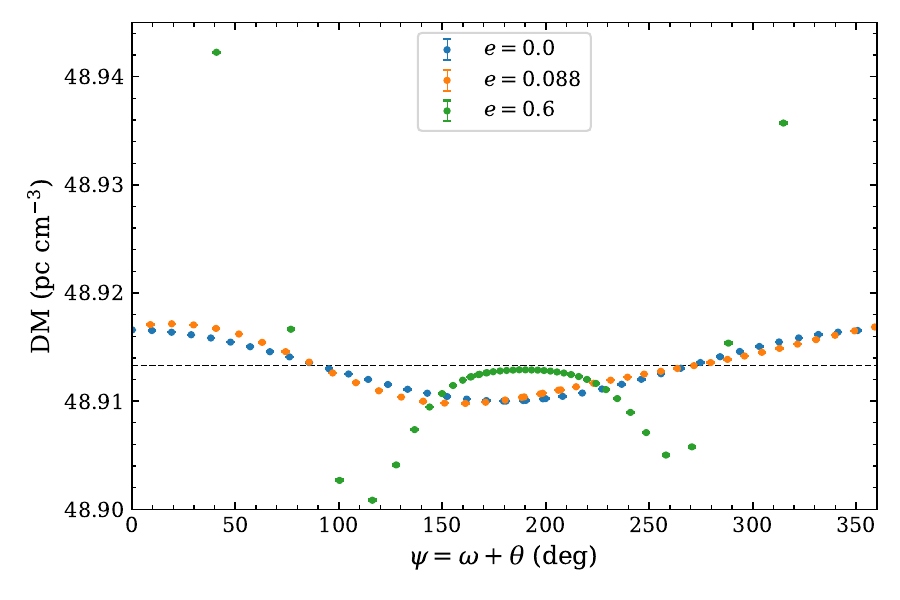}
    \vspace{-20pt}
    \caption{Apparent DM variation caused by orbital profile smearing for different eccentricities using the J0737$-$3039A profile.}
    \label{fig:dm_psrchive}
    \vspace{-10pt}
\end{figure}

In Fig.~\ref{fig:dm_psrchive} we show the apparent DM variation caused by orbital profile smearing using J0737$-$3039A's profile for different orbital eccentricities. 
A circular orbit leads to a sinusoidal variation (blue), whereas the Double Pulsar orbit ($e=0.088$) results in a slightly asymmetric variation, which is consistent with what has been seen with MeerKAT when processing the data in a similar fashion. 
The amplitude of this effect is an order of magnitude smaller than the DDV described in Section~\ref{sec:psr_doppler} and found in \cite{Ransom+2004}.
As the eccentricity increases, the variation becomes more dramatic in amplitude, as indicated by the green points in Fig.~\ref{fig:dm_psrchive} for $e=0.6$. 

To summarise, for highly relativistic binaries, observations with a sufficiently wide band could result in orbital phase-dependent profile smearing that can cause apparent phase-dependent DM variations, if TDDS is not taken into account. 
Such smearing of a profile can be eliminated if the data are de-dispersed first before integration.

\subsubsection{Spectral Dispersive Doppler Smearing}
\label{sec:psr_spectral_smearing}

In contrast to TDDS, the SDDS effect arises when spectral variation of the Doppler-shifted pulsar spin frequency across the observed bandwidth is neglected when computing and correcting the dispersive phase shift.  The sign and magnitude of the effect depends on orbital phase and is independent of integration length.

For highly relativistic systems such as the Double Pulsar, due to a significant change in the position of the pulsar in its orbit during the dispersive delay time, pulses which are received simultaneously at different frequencies were originally emitted at different orbital phases.
However, the current implementation of \textsc{psrchive} uses phase predictions at a single central frequency. 
When this predictor is used to fold data at other frequencies, it effectively neglects the difference in apparent spin frequency between different bands and causes profile smearing after averaging in frequency. 
This effect was commonly misaccepted by the community\footnote{See again comments uploaded with \cite{Ransom+2004}.} as an explanation for the DM variations seen in \cite{Ransom+2004}.

\begin{figure}[t]
    \vspace{-10pt}
    \centering
    \includegraphics[width=\columnwidth]{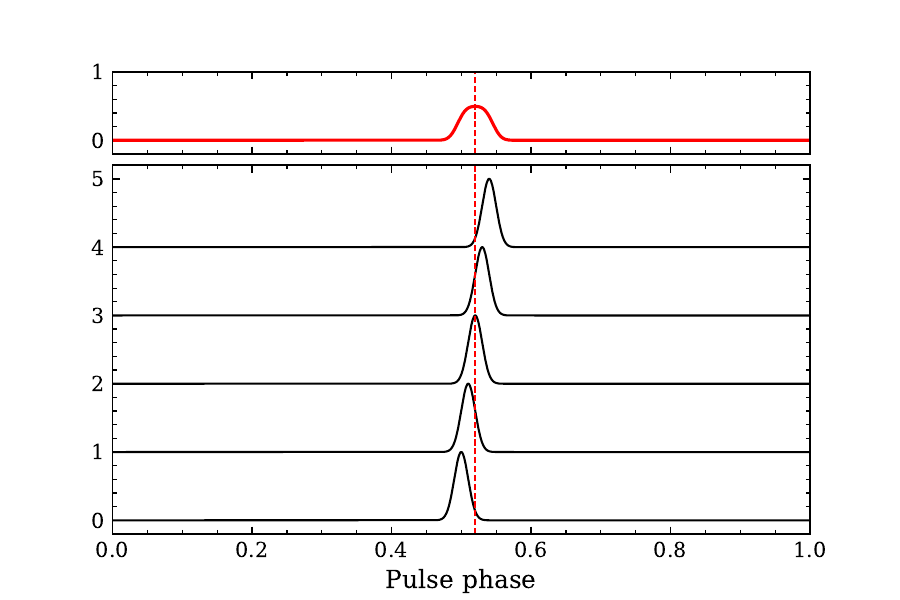}
    \vspace{-10pt}
    \caption{Schematic of folding with wrong spin periods, the resulted averaged profile (red) becomes broader and flatter. Viewing from top down, folding with a long spin period shifts the centre of the averaged profile to the left; from bottom up, folding with a short spin period shifts the averaged profile to the right.}
    \vspace{-10pt}
    \label{fig:folding}
\end{figure}

More specifically, if the data are folded with a slightly higher spin frequency (i.e. shorter spin period), the profile shifts to the right, and vice versa. 
A schematic of pulse broadening due to folding with wrong spin periods is demonstrated in Fig.~\ref{fig:folding}.
At the conjunctions ($\psi=90\si{\degree}/270\si{\degree}$), when individual profiles move both left- and rightwards, the average profile becomes broadened on both sides and the impact on the arrival time is minimal. While at the ascending and descending nodes ($\psi=0\si{\degree}/180\si{\degree}$), the profiles shift towards one direction and therefore the impact on the arrival time is maximal. 
As the effect is most prominent at low frequencies, it is covariant with scattering caused by turbulent IISM plasma. 

The spectral smearing expansion in Eq.~(\ref{eq:delta_phi}) is dominated by the leading $\ddot\phi_0$ term;
therefore, the SDDS is maximal at conjunction, where
\begin{equation}
\delta\phi_{3,\mathrm{max}}(f) \simeq \frac{a_r}{2c} \nu_0 \,
  \Delta t_\mathrm{DM}^2(f,f_0) .
\label{eq:spectral}
\end{equation}
When using a polynomial computed for $f_0=815$~MHz to fold a signal at $f=570$~MHz, the maximal inaccuracy is $\sim 37$~ns for PSR~J0737-3039A. Compared to TDDS, the SDDS is not related to integration time, but, because it is proportional to $\Delta t_\mathrm{DM}^2$, it becomes much more significant at lower radio frequency. 
For example, even though this effect is three orders of magnitude smaller than the temporal smearing at the UHF band and does not affect the timing precision, it becomes the dominant effect at low frequencies, as shown in Section~\ref{sec:ska_low}.

From a technical point of view, the SDDS can also be avoided by applying the de-dispersion routine prior to averaging. 
In this case, data at all observational frequencies effectively correspond to the same orbital phase so that the same spin frequency can be applied. 

\subsubsection{Simulations for the SKA and separating the two effects at low frequencies}
\label{sec:ska_low}

Both SDDS and TDDS lead to a frequency dependent shift that can manifest itself as orbital DM variations (especially for broad pulse profiles). Therefore, the two effects are highly covariant and difficult to disentangle. The most striking difference is evident from the Eqs.~\eqref{eq:temporal_max} and \eqref{eq:spectral}: in contrast to spectral smearing, temporal smearing depends on integration time and, therefore, there are some regimes where one is dominant over the other.

\begin{figure}[t]
    \centering
    \includegraphics[width=\columnwidth]{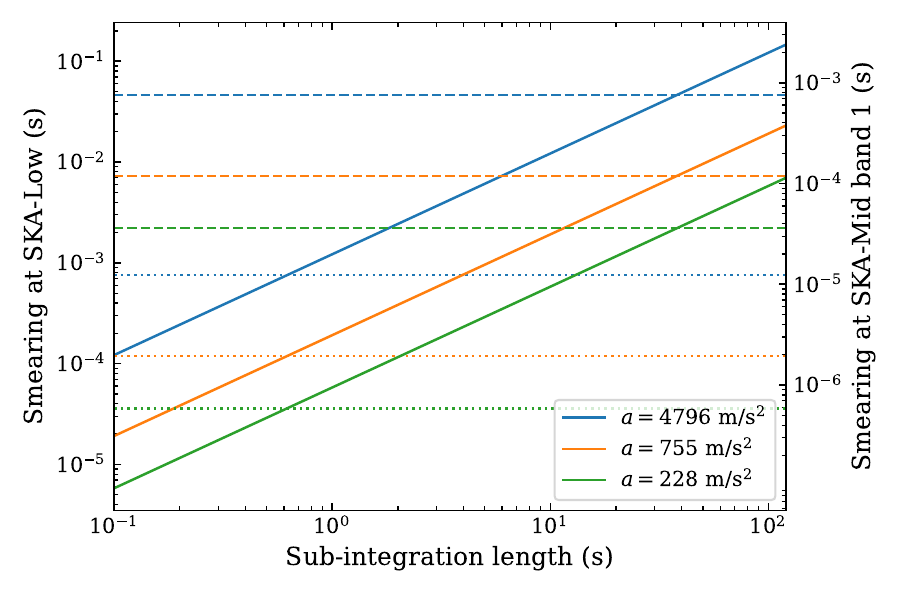}
    \vspace{-20pt}
    \caption{Maximum smearing with respect to sub-integration length for SKA-Low (left y-axis) and SKA-Mid (right y-axis) calculated using Eqs.~\eqref{eq:temporal_max} and \eqref{eq:spectral}, assuming the spin period and DM of PSR~J0737$-$3039A, mass of $1.4\, \mathrm{M}_\odot$, and circular orbits. 
    The blue, orange, and green lines denote orbital periods of 15~min, 1~h, and 2.45~h (PSR~J0737$-$3039A), respectively, with the maximum acceleration labelled. 
    The SDDS (dashed lines for SKA-Low and dotted lines for SKA-Mid) dominates over the TDDS (solid lines) when the sub-integration length is smaller than half of the dispersive delay time ($\Delta t_\mathrm{DM}=76.1$~s for SKA-Low and $\Delta t_\mathrm{DM}=1.2$~s for SKA-Mid), and vice versa. }
    \label{fig:acc}
    \vspace{-10pt}
\end{figure}

We plot the maximum magnitude of these two effects (over the orbital phase) in Fig.~\ref{fig:acc} for binaries with different orbital periods (i.e. accelerations). The left axis shows the smearing for the case of SKA-Low, whereas the right axis shows the values for the SKA-Mid, which is $\sim$60 times smaller. 
The smearing increases significantly with acceleration (for both effects) and sub-integration length (for temporal smearing). The magnitude of these two effects is approximately equal when the sub-integration length $T$ approaches half of the dispersive delay time. At SKA-Mid, the induced timing residuals are dominated by the temporal smearing starting from $T=0.6$~s, as well as for the UHF band as showed in Sections~\ref{sec:psr_temporal_smearing} and \ref{sec:psr_spectral_smearing}. However, at SKA-Low, where the dispersive delay is much longer, the smearing is dominated by the spectral Doppler effect when $T<38$~s.
\begin{figure}[t]
    \centering
    \includegraphics[width=\columnwidth]{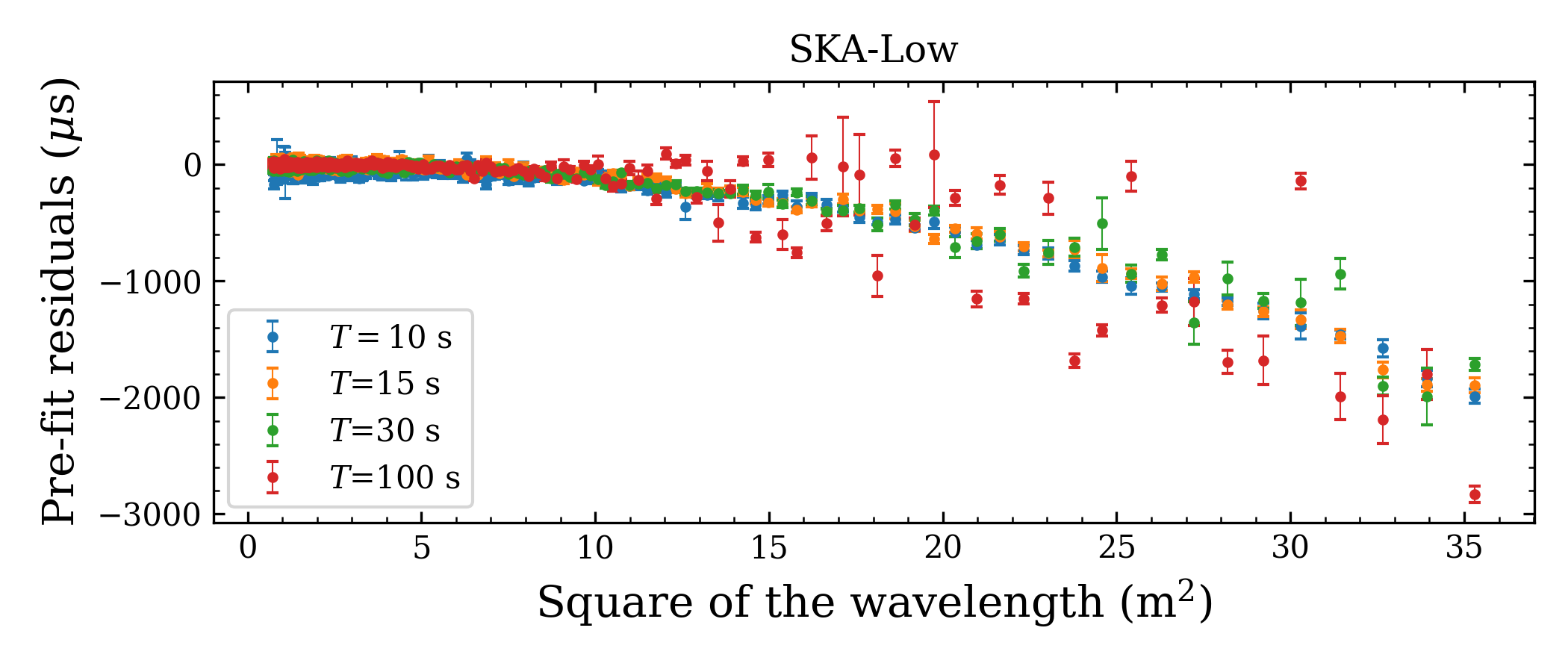}
    \includegraphics[width=\columnwidth]{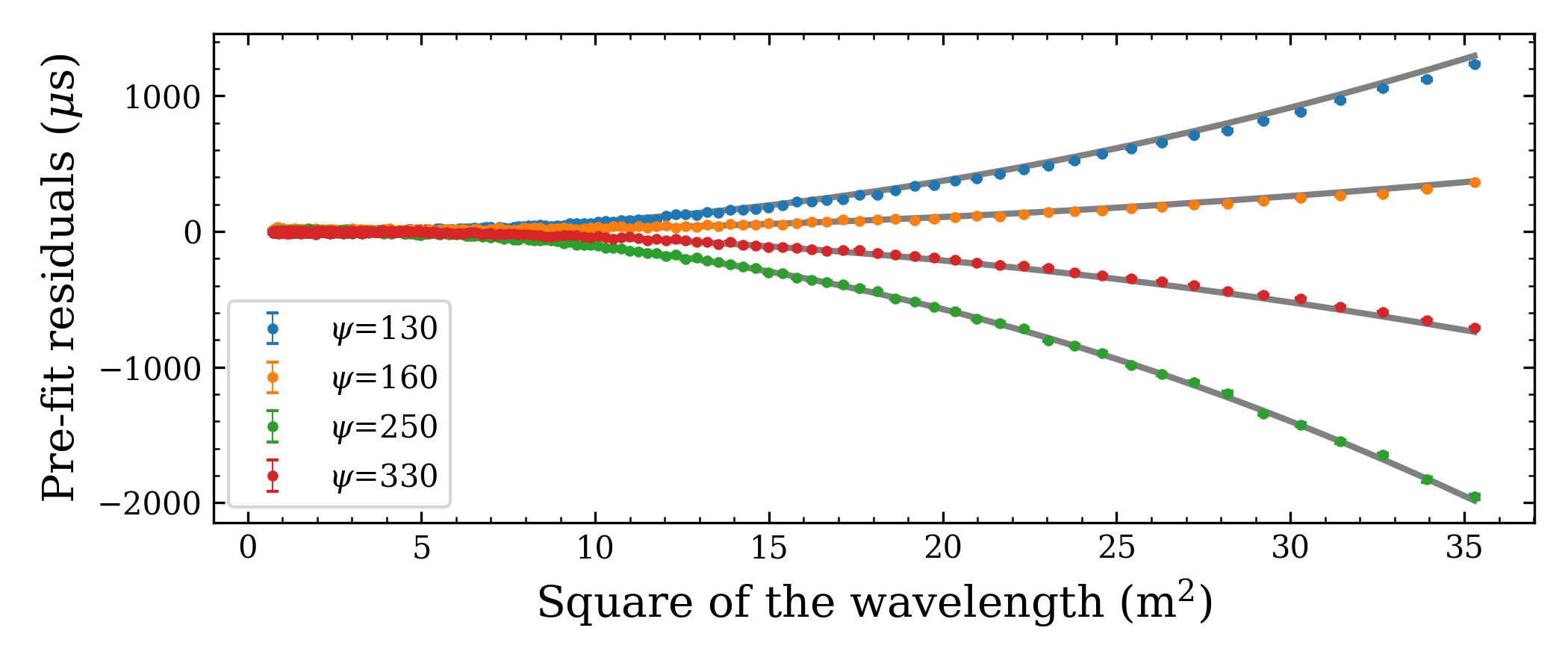}
    \vspace{-15pt}
    \caption{Upper panel: Timing residuals for a chosen epoch perturbed by SDDS and TDDS in simulated SKA-Low data for different sub-integration time $T$. 
    For $T=10,15,30$~s, the effect of SDDS prevails the TDDS and the residuals are consistent; while for $T=100$~s, the TDDS dominates and distorts the signal at lower part of the frequency band (see Fig.~\ref{fig:temp_smear2}) and increases the residuals. 
    Lower panel: Timing residuals for $T=10$~s dominated by the SDDS. Multiple colors correspond to different orbital phases $\psi$. Grey lines illustrate the theoretical magnitude of the SDDS predicted by Eq.~(\ref{eq:spectral}).}
    \label{fig:spect_smear1}
    \vspace{-5pt}
\end{figure}

\begin{figure}[t]
    \centering
    \vspace{-10pt}
    \includegraphics[width=\columnwidth]{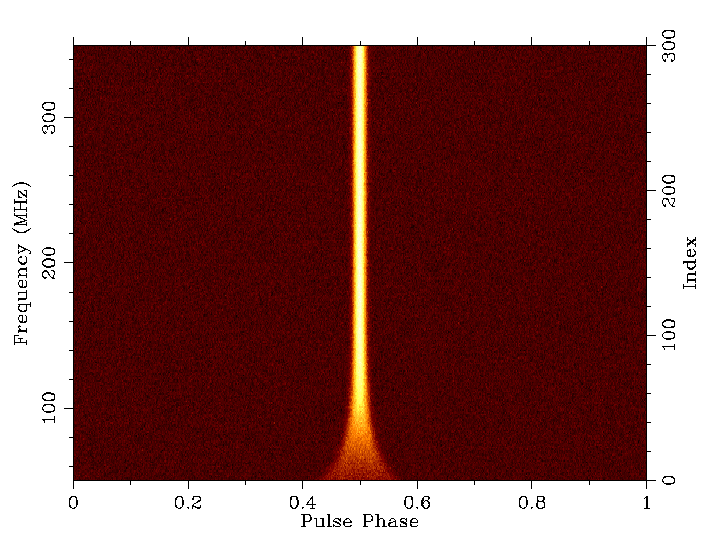}
    \vspace{-20pt}
    \caption{The TDDS effect in the simulated SKA-Low data for $T=100$~s to make the effect more prominent. The corresponding residuals are demonstrated in Fig.~\ref{fig:spect_smear1}.}
    \label{fig:temp_smear2}
    \vspace{-10pt}
\end{figure}

In order to separate these two effects and illustrate the SDDS, we simulate SKA-Low data (50--350 MHz, see Setting 2b in Table~\ref{table:detail}) with different sub-integration lengths. 
The comparison of timing residuals with different $T$ is shown in Fig.~\ref{fig:spect_smear1}. 
For sub-integration time ($T=10,15,30$~s) smaller than half of the dispersive delay time (38~s), the magnitude of residuals remains the same, indicating that the smearing is dominated by SDDS. 
The lower panel shows the timing residuals for $T=10$~s for four different orbital phases, which agrees with the theoretical prediction of Eq.~\eqref{eq:spectral} for SDDS (grey lines). 
In contrast, when the sub-integration time is much longer than half of the dispersive delay time, i.e. $T=100$~s, the magnitude of residuals increases and spreads around the mean. This different observing manifestation comes from the TDDS. 
In Fig.~\ref{fig:temp_smear2} we show the spectrum of the data for the latter case where the pulse is smeared in both directions at the lowest frequencies, resembling the ``Eiffel Tower''. 

\begin{figure}[t]
    \centering
    \includegraphics[width=\columnwidth]{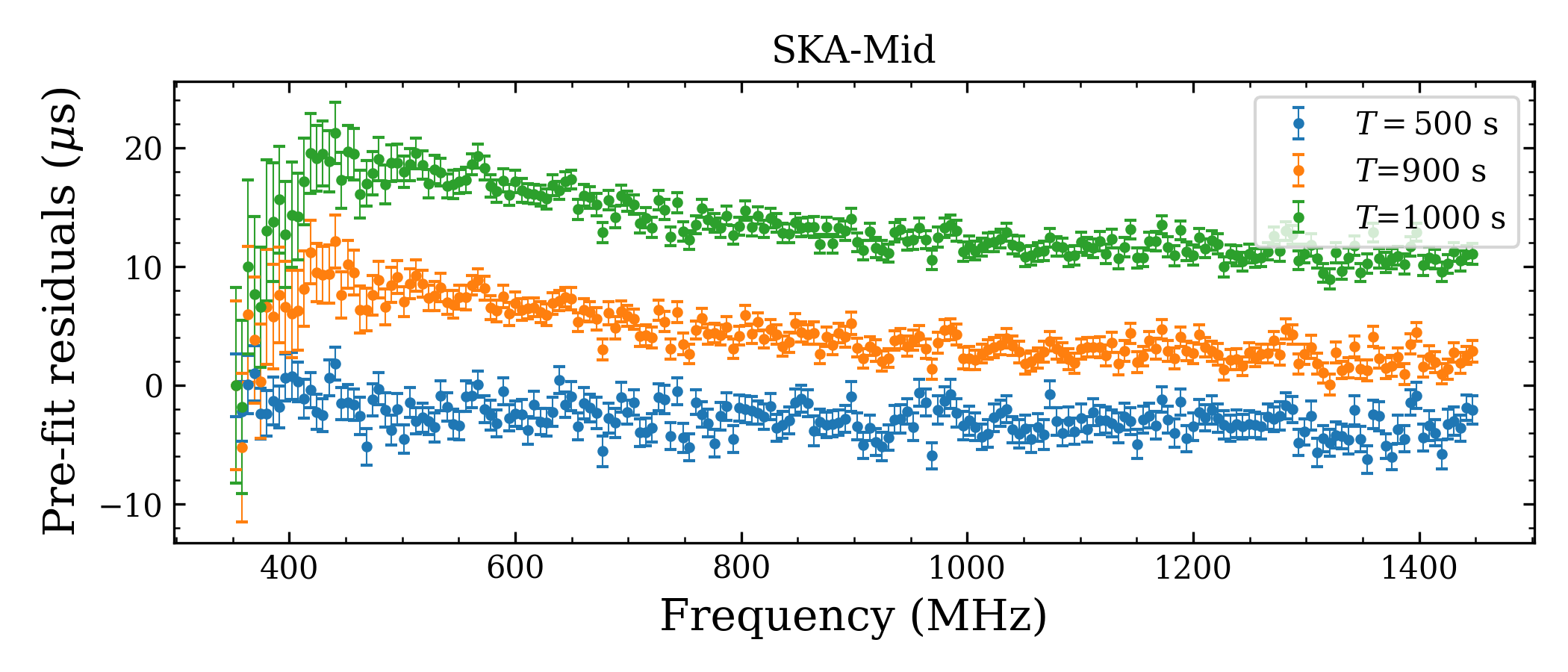}
    \vspace{-15pt}
    \caption{Timing residuals of the TDDS effect simulated for the SKA-Mid data (Setting 2b). The magnitude of the effect gradually increases with the sub-integration time $T$ as predicted by Eq.~\eqref{eq:temporal_max}.}
    \label{fig:temp_smear1}
    \vspace{-5pt}
\end{figure}

At higher frequency such as SKA-Mid (350--1450~MHz, see Setting 2c in Table~\ref{table:detail}), the effect is dominated by TDDS and is demonstrated in Fig.~\ref{fig:temp_smear1} with sub-integration time of 500~s, 900~s and 1000~s, respectively. 
The timing residuals caused by TDDS increase with sub-integration length at the given orbital phase, with the most prominent effect occurring at the lower part of the band as predicted by Eq.~(\ref{eq:temporal_max}).

\section{Topocentric Doppler effects}
\label{sec:topo_doppler}

In the previous section we have discussed three dispersive Doppler artefacts related to pulsar's orbital motion; here we complete this discussion by considering also the observatory's motion relative to the SSB and its related impact.

\subsection{Topocentric Dispersive Doppler Variation (TopoDDV)}
\label{sec:topoDDV}
In Eqs.~\eqref{eq:phi_approx} and ~\eqref{eq:phi_exact}, time and frequency are specified in the topocentric reference frame of the observatory. 
However, dispersion takes place in the rest frame of the IISM, which assumed to have constant velocity with respect to the SSB. 
Therefore, before computing the (topocentric) dispersive delay $\Delta t_\mathrm{DM}$, topocentric frequencies are first converted to barycentric frequencies, then the resulting barycentric delay is converted to a topocentric delay as follows.
Given the velocity of the observatory with respect to the SSB projected onto the line of sight to the pulsar,
$v_\mathrm{topo}(t)$, topocentric frequency $f$ is related to barycentric frequency $f_\mathrm{SSB}$ by
$f=[1+\beta_\mathrm{topo}(t)] f_\mathrm{SSB}$, where $\beta_\mathrm{topo}(t)=v_{\mathrm{topo}}(t)/c$
is positive when the observatory is moving toward the pulsar. 
Similarly, any signal propagation interval (such as wave period or dispersive delay) is transformed from the rest frame of the SSB to that of the observatory by $\Delta t = \Delta t_\mathrm{SSB}\, [1+\beta_\mathrm{topo}(t)]^{-1}$. Combining the transformations of radio frequency and time interval yields the following expression for the topocentric DDV (TopoDDV), corrected for the motion of the observatory with respect to the SSB,
\begin{equation}
\begin{aligned}
    \Delta t_\mathrm{DM}(t,f,f_0) &= \Delta t_\mathrm{DM,SSB}\left(f_\mathrm{SSB},f_\mathrm{0,SSB}\right)\, \left[1+\beta_\mathrm{topo}(t)\right]^{-1} \\
    & = D  \left[f_\mathrm{SSB}^{-2} - f_\mathrm{0,SSB}^{-2}\right] \, [1+\beta_\mathrm{topo}(t)]^{-1} \\ 
    & = D  \left(f^{-2} - f_0^{-2}\right) \,[1+\beta_\mathrm{topo}(t)] \,,\label{eq:bary}
\end{aligned}
\end{equation}
where $D=\mathcal{D}\times\mathrm{DM}$.  In the current version of {\sc psrchive}, 
the barycentric correction is disabled by default, and can be enabled by setting
{\tt Dispersion::barycentric\_correction = 1} in the {\tt psrchive.cfg} file.
When enabled, the correction is approximated by 
\begin{equation}
    \Delta t_\mathrm{DM}(t,f,f_0) \simeq D  \left(f^{-2} - f_0^{-2}\right) \, [1-\beta_\mathrm{topo}(t)]^{-1},
\end{equation}
which is valid for small $\beta_\mathrm{topo}(t)$.
If this correction is not enabled, Earth's orbital velocity will induce an apparent annual modulation 
of DM with a maximum relative amplitude of $\delta\DM_\mathrm{max}/\DM \simeq 10^{-4}$ for pulsars near the ecliptic plane.  A much smaller apparent diurnal variation of DM due to Earth's rotation has a maximum 
relative amplitude of $\delta\DM_\mathrm{max}/\DM \simeq 10^{-6}$ for a pulsar and observatory near the equator.

\subsection{Topocentric Temporal Dispersive Doppler Smearing (TopoTDDS)}
\label{sec:earth_smearing}

When phase is predicted using Eq.~\eqref{eq:phi_approx}, the variation of the velocity 
of the observatory over the sub-integration time also leads to phase drift 
and profile smearing, as demonstrated in Section 7 of HEM2006. Including this variation in Eq.~\eqref{eq:delta_phi} gives rise to an additional term
\begin{equation}
\begin{aligned}
\label{eq:delta_phi_topo}
\Delta\phi_\mathrm{topo}(t,f) 
&= \dot\phi_0 \left[ \Delta t_\mathrm{DM}(t,f,f_0) - \Delta t_\mathrm{DM}(t_\mathrm{mid},f,f_0) \right] \\
&= \nu_0 D \left(f^{-2} - f_0^{-2}\right) \left[ \beta_\mathrm{topo}(t) - \beta_\mathrm{topo}(t_\mathrm{mid}) \right].
\end{aligned}
\end{equation}
Over the duration of the sub-integration $T$, the total smearing is, to the first order,
\begin{align}
    \delta \phi_\mathrm{topo}(t,f) \simeq \frac{a_\mathrm{topo}}{c} T \Delta t_\mathrm{DM} (f, f_0) \, \nu_0 \,,
\end{align}
Where $a_\mathrm{topo}$ is the acceleration of the observatory with respect to the SSB, projected onto the line of sight to the pulsar.
For a pulsar in the ecliptic plane with DM=50~$\mathrm{pc \, cm^{-3}}$, integrating for $T=30$~s using a polynomial computed 
for $f_0=200$~MHz to fold a signal at $f=50$~MHz results in maximum smearing of $\sim 46$~ns.
In MeerKAT L-band observations of PSR~J0737$-$3039A, the effect is two orders of magnitude smaller.

\section{Folding issues with phase predictors}
\label{sec:folding}
\begin{figure}[t]
    \centering
    \includegraphics[width=0.9\columnwidth]{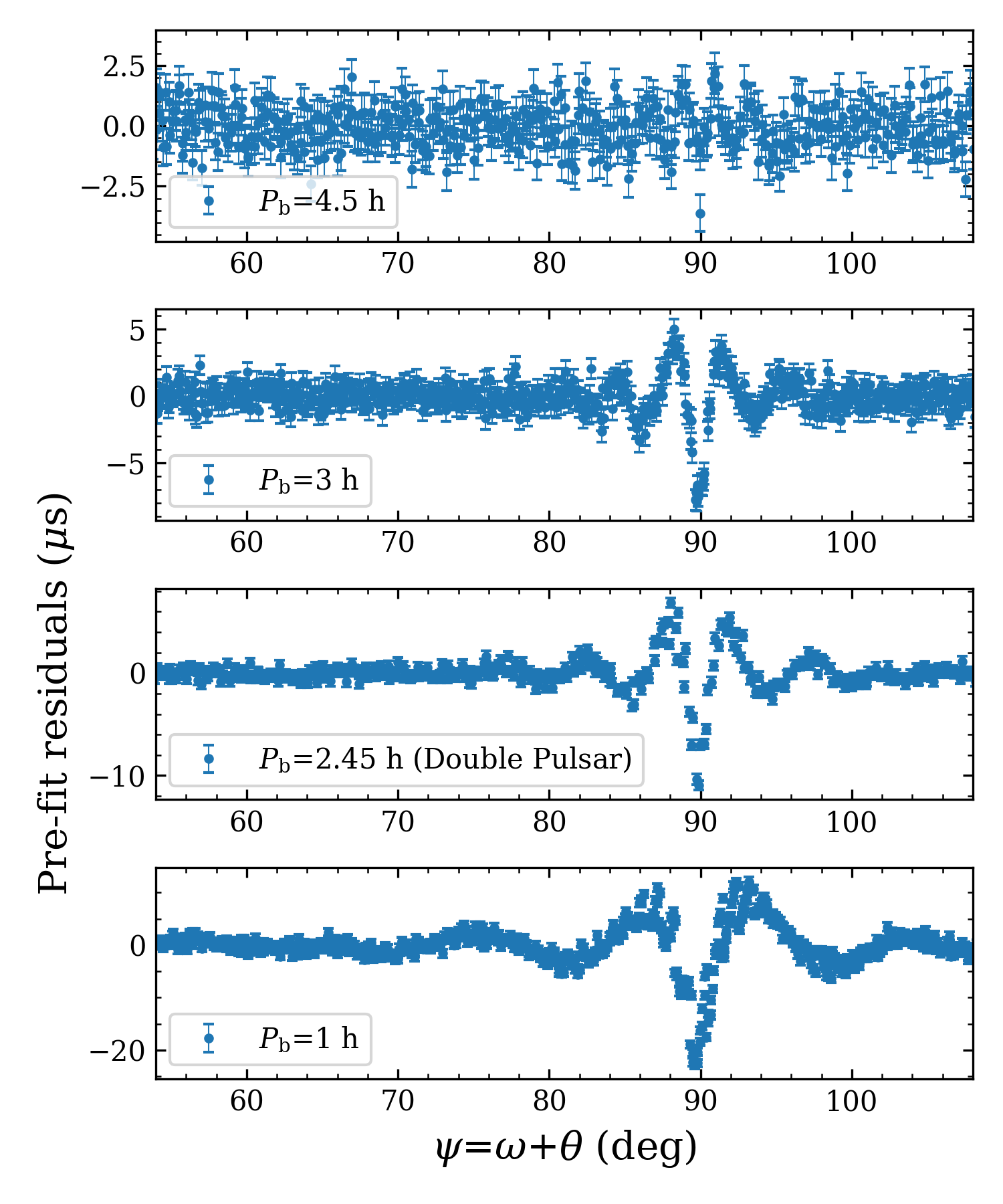}
    \vspace{-10pt}
    \caption{TOA residuals of the simulated data files vs. orbital phase for binary systems with orbital periods of 4.5~h, 3~h, 2.45~h and 1~h. The spiky features appearing at the superior conjunction ($\psi=90\si{\degree}$) demonstrate the limitations of the current software, in this case, the ``polyco'' predictor. The simulation was done with TZNSPAN=3~min and 20 coefficients. }
    \label{fig:dft}
\end{figure}

\begin{figure}[t]
    \vspace{-10pt}
    \centering
    \includegraphics[width=0.9\columnwidth]{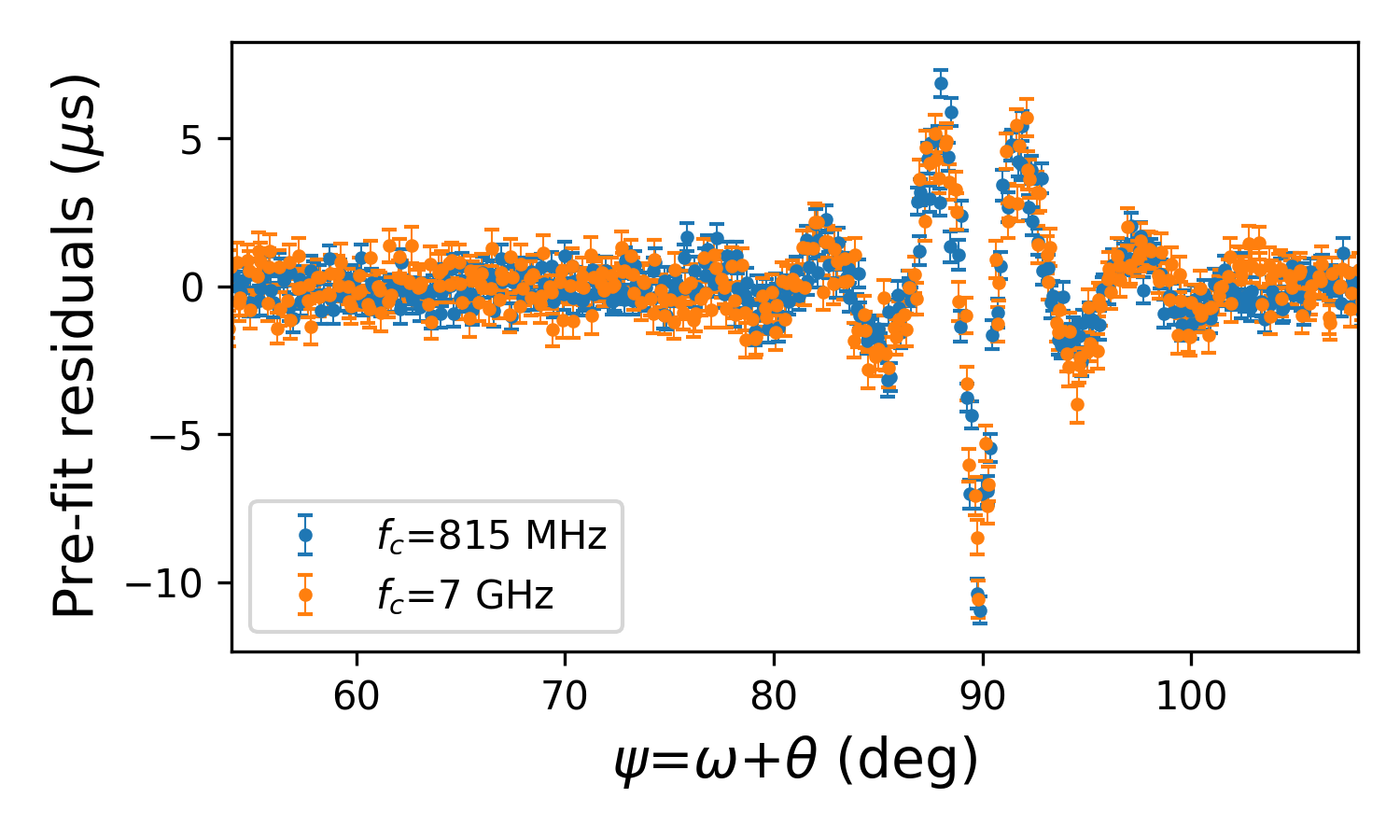}
    \vspace{-10pt}
    \caption{Same as Fig.~\ref{fig:dft} but for the Double Pulsar with observations simulated at two remote radio frequencies of 815~MHz (blue) and 7000 MHz (orange). The figure clearly shows that the systematic issue due to folding is independent of observational frequency.}
    \label{fig:dft_high_low}
    \vspace{-10pt}
\end{figure}
\begin{figure*}[t]
    \centering
    \vspace{-10pt}\includegraphics[width=\textwidth]{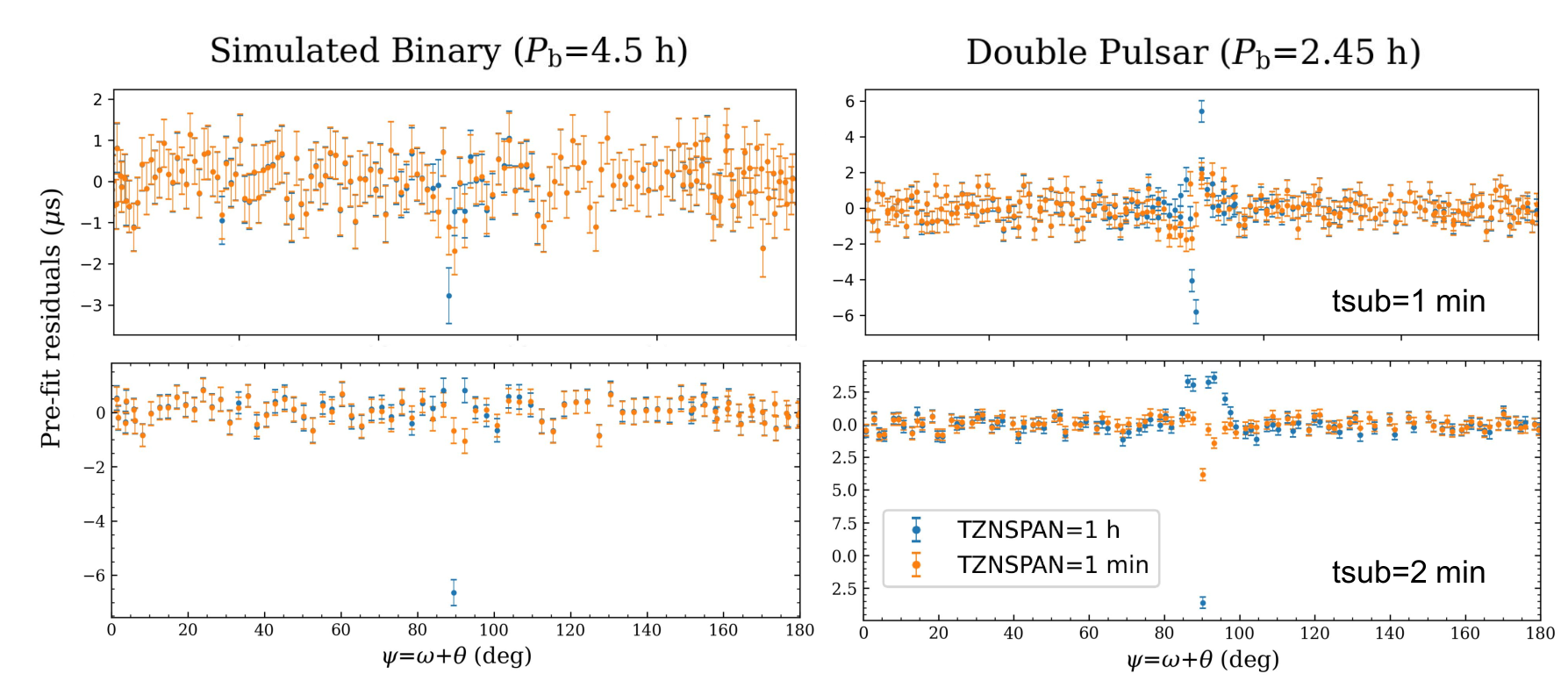}
    \vspace{-20pt}
    \caption{TOA residuals of the simulated data files as a function of the orbital phase: for a fictional binary with orbital period $P_\mathrm{b}=2.45$~h on the left and for the Double Pulsar on the right. For all cases the original time resolution of the generated data files is TZNSPAN=1~min. Results with the originally installed 1-min resolution ``polyco'' predictor are in orange, while in blue colour the results with re-installed ``polyco'' with TZNSPAN=1~h resolution are shown. The upper panel shows the residuals after folding the originally generated fits files and reducing the time resolution by a factor of 60 (so that the resultant time resolution of the files is 1 min). The bottom panel is the same, but the time-averaging factor is 120 (2 min). The spiky features at phase $\sim0.25$ represent limitations of the ``polyco'' predictor, and become more pronounced as the time resolution of the predictor deteriorates and the sub-integration time (``tsub'') increases. The effect is amplified for more relativistic binaries.}
    \label{fig:dft_factor}
\end{figure*}

Another systematic issue that arises when timing highly relativistic pulsar binaries is due to the limited accuracy of current phase predictors. 
When averaging the data in frequency or time, \textsc{psrchive} and \textsc{dspsr} do not use the pulsar ephemeris directly, but rather use pulse phase predictions.
Commonly, the pulse phase prediction information is supplied by a predictive polynomial file derived from the pulsar ephemeris, which are ``polyco'' or ``T2predict'' files generated using the pulsar timing software \textsc{tempo} and \textsc{tempo2}, respectively. 
This contains a set of polynomial coefficients that describe the pulse phase and spin frequency for a given observation frequency and time range. The parameter TZNSPAN defines the valid time span covered by each unique set of coefficients.
One should emphasise that the predictor does not represent the exact timing solution, but rather the approximation of it using polynomial/Chebyshev expansion.
These approximated polynomials generally describe the timing solution well, but fail for relativistic binaries especially at the superior conjunction due to the sharpness of the Shapiro delay, which follows a natural logarithm \citep{BT1976,DD86}:
\begin{align}
    \Delta_\mathrm{S}^\mathrm{(LO)} &= -2r \ln{\Lambda_u} \,,\\ 
   \Lambda_u & = 1-e \cos{u} - s\,[\sin{\omega}\,(\cos{u}-e) \notag\\ 
   & \quad + (1-e^2)^{1/2} \cos{\omega}\sin{u}] \,. \label{eq:cp2shapiro}
\end{align}
The parameters $r$ and $s$ are the post-Keplerian parameters representing the ``range'' and ``shape'' of the Shapiro time delay as the pulsar signal passes in the vicinity of the companion, which can be measured using pulsar timing. $e$ is the orbital eccentricity, $u$ is the relativistic eccentric anomaly, and $\omega$ is the longitude of periastron.
The limitation of polynomial approximation has been known for a long time, which can in principle affect all timing parameters, but is most easily seen in the Shapiro delay parameters.
As noted by \cite{Hu+2022}, inappropriate choice of a predictor and relevant parameters (TZNSPAN and the number of polynomial coefficients---``TZNCOEF'') can cause bias in the measurements of Shapiro delay parameters.

In order to demonstrate the difference between the exact timing solution and the ``polyco''-predicted phase, we simulate a series of data files with the setup listed in Table~\ref{table:detail}. 
For comparison purposes, we generate pulsar data files with different orbital periods, and here we choose to present four cases: 4.5~h, 3~h, 2.45~h (i.e. the Double Pulsar) and 1~h. 
For all four cases the resolution of the predictor, defined by the TZNSPAN parameter, is fixed to 3 min. After generation, fully resolved data files are de-dispersed and frequency-averaged. 
The pulsar data files are generated with ultra-high signal-to-noise ratio, so their precision exceeds that of data from existing telescopes.
Given that \textsc{PsrSigSim} generates pulsar data files based on the polynomial expansion of the timing solution (predictor), the discrepancies of interest should be visible in the residuals after subtracting the best-fit timing model and are indeed shown in Fig.~\ref{fig:dft}. As expected the difference is most prominent at orbital phase $\psi=90\si{\degree}$, which corresponds to the superior conjunction of the binary system. Fig.~\ref{fig:dft} also clearly shows that the systematic error increases as the orbital period decreases. One should note here that the considered effect does not depend on observational frequency and bandwidth, as it is demonstrated in Fig.~\ref{fig:dft_high_low}.

Such discrepancies are not immediately visible in real observations, but will propagate in the timing analysis at specific stages of post-processing. To be more precise, systematic error arises at the superior conjunction of highly relativistic binaries, when forming the sub-integrated data with, e.g. \textsc{dspsr} software, which uses polynomial expansion in its base. 
To simulate the effect as it appears in the real data, we create pulsar data files with 1~s sub-integration length and 1~min time resolution for the predictor (defined by the parameter TZNSPAN), which is the smallest possible TZNSPAN that is compatible with the \textsc{psrchive} software. 
This allows us to mimic realistic observations as closely as possible.
Data were simulated for PSR~J0737$-$3039A and a fake binary pulsar with an orbital period of 4.5~h (Setting 3b in Table~\ref{table:detail}).
When averaging the data only in frequency with TZNSPAN=1~min, the systematic error of interest is not visible in either system. 
However, if the resolution of the ``polyco'' predictor in the original file is reduced (i.e. a larger TZNSPAN) and the sub-integration length of the data is increased, the systematic effect appears.

{\renewcommand{\arraystretch}{1.4}
\begin{table}[t]\small
\caption{Estimated bias in Shapiro parameters $s$ and $r$ for MeerKAT and the SKA based on simulated data. }
\centering
\begin{tabular}{lccc} 
\hline \hline
Telescope & MeerKAT & SKA AA* & SKA AA4 \\ \hline 
Number of dishes & 64 & 144 & 197 \\ \hline
Effective diameter (m) & 108 & 172 & 204 \\ \hline
Estimated RMS ($\mu$s) & 9.4 & 3.7 & 2.6 \\ \hline
Bias in Shapiro $s$ & 0.7 $\sigma$ & 1.8 $\sigma$ & 2.5 $\sigma$\\ \hline
Bias in Shapiro $r$ & 0.6 $\sigma$ & 1.4 $\sigma$ & 2.0 $\sigma$ \\
\hline
\end{tabular}
\tablefoot{Two stages for the SKA-Mid are shown, the AA* in 2028 with 144 dishes and AA4 with 197 dishes. For all simulations, ``polyco'' with TZNSPAN=1~min and 30 coefficients were adopted. }
\label{table:shap}
\vspace{-5pt}
\end{table}

{\renewcommand{\arraystretch}{1.3}
\begin{table*}[t]\small
\caption{List of all artefacts discussed in this paper, including their causes, manifestations and solutions. }
\centering
\begin{tabular}{m{4.5cm}|m{6.0cm}|m{6.6cm}} 

\hline \hline
\makecell{Artefact} & \makecell{Cause and Manifestation} & \makecell{Solution} \\ \hline 
\makecell{Dispersive Doppler Variation (DDV) \\ (see Section~\ref{sec:psr_doppler})}
& Neglecting the variation of pulsar spin frequency when computing the dispersive phase shift. It causes an apparent orbital DM variation, with maximum and minimum values at $\psi=180\si{\degree}$ and  $\psi=0\si{\degree}$, respectively.
& The spin frequency variation has been implemented in \textsc{psrchive} and \textsc{dspsr} and no longer results in apparent orbital DM variation, while for other software such as \textsc{sigproc}, the problem still persists. \\ \hline
\makecell{Temporal Dispersive Doppler \\ Smearing (TDDS) \\ (see Section~\ref{sec:psr_temporal_smearing})}
& 
Neglecting the variation of spin frequency over the duration of the sub-integration when computing the dispersive phase shift. It causes profile smearing that increases with sub-integration duration and results in an apparent orbital DM variation with a maximum value at $\psi=0\si{\degree}$.
& De-dispersion before integrating the data in time while using \textsc{dspsr} and \textsc{psrchive}. \\ \hline
\makecell{Spectral Dispersive Doppler \\ Smearing (SDDS)  \\ (see Section~\ref{sec:psr_spectral_smearing})}
& Neglecting the differential spin frequency over the observed bandwidth when computing the dispersive phase shift. It causes a similar profile smearing effect as TDDS; however, it does not depend on integration length.
& De-dispersion before integrating the data in time while using \textsc{dspsr} and \textsc{psrchive}, or implementing a two-dimensional phase predictor.
\\ \hline
\makecell{Topocentric Dispersive Doppler \\ Variation (TopoDDV) \\ (see Section~\ref{sec:topoDDV})}
& Neglecting the barycentric correction of observed radio frequency when computing the dispersive delay. 
It causes an apparent annual DM variation.
& This correction can be enabled in \textsc{psrchive} by setting {\tt Dispersion::barycentric\_correction = 1} in the {\tt psrchive.cfg} file. \\ \hline
\makecell{Topocentric Temporal Dispersive \\ Doppler Smearing (TopoTDDS)  \\ (see Section~\ref{sec:earth_smearing})}
& Neglecting the variation of the velocity of the observatory over the sub-integration time. It causes similar profile smearing effect as TDDS.
& The effect is negligible and can be reduced by shortening the sub-integration time. \\ \hline
\makecell{Folding issues with phase predictors \\ (see Section~\ref{sec:folding})}
& Inaccurate polynomial phase predictions at superior conjunction. It causes biases in timing parameters, most notably in the Shapiro delay parameters. 
& Decreasing the time span and increasing the number of coefficients of each set of polynomials hit the limit for the SKA. Development of a more accurate phase predictor that describes the Shapiro delay is needed. \\ \hline
\end{tabular}
\label{table:list}
\end{table*}

As demonstrated in Fig.~\ref{fig:dft_factor}, the longer the sub-integration length and the worse the resolution of the phase predictor, the larger the effect. 
Specifically, the systematic error of interest is shown for two different values of TZNSPAN, 1 min and 1 h, and two different time-averaging factors, which are 60 and 120, corresponding to 1 min and 2 min length of sub-integration. 
For the Double Pulsar, the size of the effect is as high as $\sim5\,\mu$s for 30-s sub-integration time and TZNSPAN=1~min, which is still below the root mean square (RMS) of MeerKAT data with the same sub-integration \citep{Hu+2022}. 
However, it can bias the Shapiro parameter measurements and is expected to be more visible in a few years with the SKA. 
In Table~\ref{table:shap} we estimate the expected bias in the Shapiro parameters $s$ and $r$ for MeerKAT and the SKA-Mid at different stages: the Array Assembly (AA)* in 2028 and the final AA4\footnote{See \url{https://www.skao.int/en} for updates.}. Although the accuracy of long-term post-Keplerian parameters is not studied in this work, it is expected that they will also be affected by inaccuracies in the modelling of the Shapiro delay.

Although the Double Pulsar is not in the field of view of FAST, currently the most sensitive radio telescope, other relativistic binaries such as PSR~J1946+2052 may face the same problem.
For sensitive telescopes such as FAST and the SKA, shorter integration times can be used to achieve the target sensitivity. Despite this, due to the improved sensitivity of these instruments, the folding problem will be even more evident due to the limited accuracy of the polynomial approximation and the limited implementation of the TZNSPAN parameter in the current software. 
In order to mitigate this folding artefact, the temporal resolution of the phase predictor and/or the approximation method needs to be improved.

\section{Summary and discussion}
\label{sec:summary}

In this paper, we discussed several data processing issues that could arise in compact binary pulsar systems and summarised their causes, manifestations and solutions in Table~\ref{table:list}.

Firstly, we have appropriately addressed the nature of the apparent DM variation seen in \cite{Ransom+2004}. For the first time, we have used mathematical expressions and simulations to show that the effect is due to an unaccounted first-order dispersive Doppler effect, the DDV. 
This Doppler shift of spin frequency has been implemented in the computation of dispersive phase shift in commonly-used pulsar processing software \textsc{psrchive} and \textsc{dspsr}, so such artefacts should not occur when processing the data with these tools.

Secondly, we have looked at two second-order dispersive Doppler effects that could cause orbital dependent profile smearing and mimic DM variation.
TDDS occurs when a constant spin frequency is used to correct dispersion over the duration of the sub-integration time, whereas SDDS occurs when a constant spin frequency is applied over the observed bandwidth when computing the dispersive delay.
Although the manifestations of both effects are similar, TDDS dominates over SDDS unless for observations at very low frequencies.
Therefore in Section~\ref{sec:ska_low}, we considered the special case of observing relativistic pulsar binaries using SKA-Low (50-350~MHz).
At these low radio frequencies, for a pulsar in a circular orbit with a period of 1~h, folding with a one-dimensional phase predictor will result in significant TDDS of the order of 1~ms per 5~s of integration and SDDS of the order of 10~ms.
A pulsar in an orbit with a period of 15~min will suffer TDDS of the order of 1~ms for each second of integration and SDDS of the order of 100~ms. 
Both TDDS and SDDS can be prevented, however, by de-dispersing the data prior to time integration (e.g. removing the inter-channel dispersive delays with ``-K'' option in \textsc{dspsr}). In addition, SDDS can also be avoided by utilising a two-dimensional phase predictor (or equivalent).

Finally, we have shown that the current folding techniques with polynomial approximation fail to describe the Shapiro delay with sufficient accuracy for compact binaries with < 4~h edge-on orbit and could cause problems in the measurements of Shapiro delay parameters. 
As Shapiro delay parameters are often measured with high precision in a short timescale compared to other post-Keplerian parameters, they play important roles in testing gravity \citep{Freire2024LRR}, measuring neutron star masses and moment of inertia \citep{Hu+2020,Hu2024Univ}, and subsequently help to constrain the equation of state of matter at supranuclear densities.
This is not only relevant to the Double Pulsar, but also to other compact binaries, such as the 1.88~h-orbit double neutron star system PSR~J1946+2052. 
Ongoing pulsar surveys are expected to discover more binary pulsar systems with tighter orbits, where the limitation of polynomial approximation becomes even more prominent. 
Such previously undetectable artefacts will become more visible with the ever-improving sensitivity of the radio facilities. 
In particular, the FAST and the SKA (under construction) are expected to be significantly affected by these signatures and potentially limit the accuracy of timing parameters and consequently gravity tests. 
Therefore, a more accurate phase predictor is indispensable to minimise the systematic error in data processing and take the precision of gravity tests and the study of extreme matter to new heights. 
\begin{acknowledgements}
We thank the anonymous referee for valuable and constructive feedback.
We are grateful to Ramesh Karuppusamy for fruitful discussions and careful reading of the manuscript. 
HH acknowledges constant support from the Max Planck Society and funding by the MPG-CAS LEGACY collaboration. 
NKP is funded by the Deutsche Forschungsgemeinschaft (DFG, German Research Foundation)– Projektnummer PO 2758/1–1, through the Walter–Benjamin programme.
\end{acknowledgements}

\bibliographystyle{aa}
\bibliography{main}

\end{document}